\documentclass[aps,prl,twocolumn]{revtex4-2}
\usepackage{bm}
\usepackage{graphicx}
\usepackage{amsmath}
\usepackage{amssymb}
\usepackage[utf8]{inputenc}
\usepackage[T1]{fontenc}
\usepackage{color}
\usepackage{float}
\usepackage{cancel}
\usepackage{upgreek}
\usepackage{subfigure}
\usepackage[normalem]{ulem}
\usepackage[unicode=true,colorlinks=true,citecolor=blue]{hyperref}
\usepackage{placeins}
\usepackage{lipsum}
\usepackage{epsfig}
\usepackage{booktabs}
\usepackage[utf8]{inputenc}
\usepackage{wrapfig}
\newcommand{\nix}[1]{}

\usepackage{ulem}
\usepackage{physics}

\begin{document}

\title{Nonlinear intensity dependence of terahertz edge photocurrents in graphene}

\author{S. Candussio$^1$,  L. E. Golub$^2$, S. Bernreuter$^1$, T. Jötten$^1$, T. Rockinger$^1$, K. Watanabe$^3$, T. Taniguchi$^4$, J. Eroms$^1$, D. Weiss$^1$, and S.D. Ganichev$^1$
}

\affiliation{$^1$Terahertz Center, University of Regensburg, 93040 Regensburg, Germany}
\affiliation{$^2$Ioffe Institute, 194021	St. Petersburg, Russia}
\affiliation{$^3$Research Center for Functional Materials, National Institute of Material Science, 1-1 Namiki, Tsukuba 305-0044, Japan}
\affiliation{$^4$International Center for Materials Nanoarchitectonics, National Institute of Material Science, 1-1 Namiki, Tsukuba 305-0044, Japan}

\begin{abstract}
We report on the observation of terahertz radiation induced edge photogalvanic currents in graphene, which are nonlinear in intensity. The increase of the radiation intensities up to MW/cm$^2$ results in a complex nonlinear intensity dependence of the photocurrent. The nonlinearity is controlled by the back gate voltage, temperature and radiation frequency. A microscopic theory of the nonlinear edge photocurrent is developed. Comparison of the experimental data and theory demonstrates that the nonlinearity of the photocurrent is caused by the interplay of two mechanisms, i.e.  by direct inter-band optical transitions and Drude-like absorption. Both photocurrents saturate at high intensities, but have different intensity dependencies and saturation intensities.  
The total photocurrent shows a complex sign-alternating intensity dependence. The  functional behaviour of the saturation intensities and amplitudes of both kinds of photogalvanic currents depending on gate voltages, temperature, radiation frequency and polarization is in a good agreement with the developed theory.
\end{abstract}

\maketitle

\section{Introduction}

Edge electron transport in graphene and topological insulators exhibiting phenomena as, e.g., quantum Hall effect~\cite{Novoselov2006,Neto2009} and quantum spin Hall effect~\cite{Murakami2004,Kane2005,Bernevig2006,Koenig2007},  is at the core of the physics of these fascinating materials.
An important access to the edge transport provides the study of the photocurrents excited by polarized light in the mean-free path vicinity of the edges. Photon helicity driven chiral~\cite{Karch2011,Otteneder2020} and helical~\cite{Dantscher2017,Kiemle2021} currents excited by circularly polarized radiation, as well as currents driven by linearly polarized radiation at zero magnetic fields~\cite{GlazovGanichev_review,Magarill2015,Plank2019,Durnev2019,Ma2019,Wang2019,Candussio2020,Otteneder2020,Durnev_pssb2021,Durnev2021PRB}, quantum Hall~\cite{Plank2019} or cyclotron resonance~\cite{Candussio2021} regimes present several examples of the edge photocurrents, for reviews see Refs.~\cite{GlazovGanichev_review,Ivchenko2018,Durnev2019}. The edge photocurrents are typically excited by terahertz or infrared radiation and so far studied only at rather low radiation intensity $I$, at which the current scales linearly with $I$, i.e., is proportional to the square of the radiation electric field $E(\omega)$. 

\begin{figure}
	\centering
	\includegraphics[]{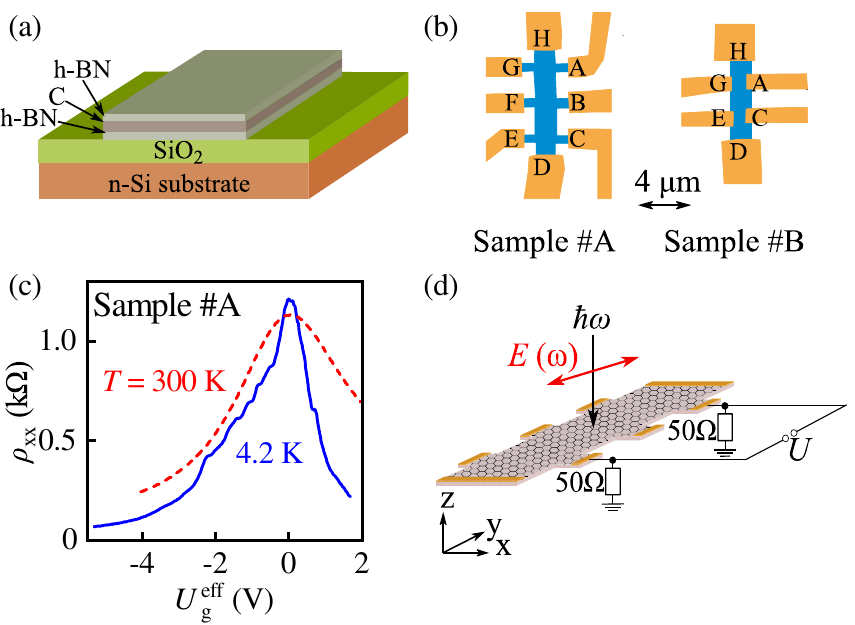}
	\caption{(a) Cross section of the graphene device, displaying the exfoliated graphene sandwiched between boron nitride.  (b) Sketch of the Hall bar shaped samples \#A and \#B with corresponding contact labels. Panel (c) shows the longitudinal resistivity $\rho_{xx}$ as a function of the effective gate voltage, obtained for room and liquid helium temperatures. (d) Schematic representation of the experimental setup. The photocurrent is measured as a voltage drop over 50 $\Omega$ load resistors.
	}
	\label{fig_1}
\end{figure}

Here we report the observation and detailed study of the nonlinear edge photocurrents excited in monolayer graphene by high power linearly polarized terahertz radiation. We show that depending on the back gate voltage and temperature the photocurrent either saturates with an increase of the radiation intensity, or, shows a non-monotonic intensity dependence, for which the current first gets saturated but at higher intensities flips its sign. Our analysis demonstrates that the latter is caused by the interplay of edge photocurrents stemming from indirect optical transitions within the conduction or valence band (Drude-like absorption) and direct inter-band optical transitions exhibiting different behaviour with the increase of the radiation intensity.  In both cases the photocurrents arise due to the momentum alignment caused by polarized radiation and asymmetric scattering in the mean-free path vicinity of the edges. The saturation of the former photocurrent is due to electron gas heating resulting in the saturation of the absorbance coefficient, previously studied applying nonlinear ultrafast THz spectroscopy~\cite{Mics2015}. The saturation of the edge photocurrent due to the inter-band transitions is caused by slow relaxation of photoexcited carriers and is described by a different power-law. While the Drude edge photocurrent has opposite polarity 
for electron and hole conductivity, the sign of the inter-band photocurrent, resulting from the generation of electron-hole pairs, does not depend on the carrier polarity.
This fact, together with the different power laws of these photocurrents, causes the sign inversion with increasing power detected at low positive back gate voltages. We  develop a theory of the nonlinear photocurrent excited by the inter-band transitions, which describes the experimental findings well. The analysis of the observed non-linear photocurrents give important information on the mechanisms of saturation and the behavior of the saturation parameters upon variation of the gate voltages, temperature, and radiation frequency. The obtained characteristics and the functional behavior of THz-radiation saturation is important for the development of terahertz mode-locking laser systems applying graphene as a material for the saturable absorbers, which are already widely used in the infrared/visible spectral ranges~\cite{Bao2009,Kumar2009,Vasko2010,Weis2012,Avouris2012,Winnerl2013,Michailov2017,Bianchi2017,Autere2018,Baudisch2018,Yumoto2018,You2019,Raab2019,Kovalev2021,Mittendorff2021}.

\begin{figure}
	\centering
	\includegraphics[]{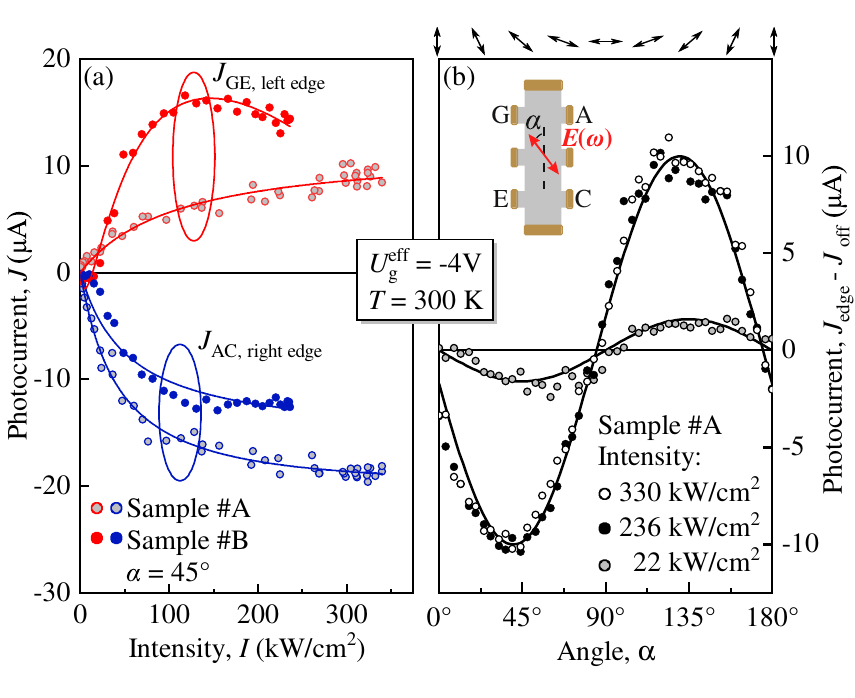}
	\caption{ (a) Intensity dependencies of the photocurrents obtained in samples \#A and \#B for a pair of contacts at the left and right sample edges. All curves are obtained for an angle $\alpha = 45^\circ$. Curves are fits to Eq.~(\ref{both}).  (b) Polarization dependence of the edge photocurrent $J_{\rm edge} = (J_{\rm AC}-J_{\rm GE})/2$ obtained for different radiation intensities. Note that a small polarization independent offset was subtracted for visibility. Curves are fits to Eq.~(\ref{eq1}). The inset defines the angle $\alpha$ describing the relative orientation of the radiation electric field vector $\bm E$ in respect to the samples long side. Arrows on top 	illustrate the polarization plane orientation for several angles $\alpha$. 
	}
	\label{fig_2}
\end{figure}

\section{Samples and methods}
\label{samples_methods}

The experiments are carried out on several Hall bar structures prepared from exfoliated graphene/hexagonal boron nitride stacks~\cite{Dean2010,Wang2013,Sandner2015}, see Fig.\,\ref{fig_1}(a) and (b). 
Sweeping the back gate voltage shows that the charge neutrality point (CNP) is well observable in the longitudinal resistance for both room and liquid helium temperatures, see Fig.~\ref{fig_1}(c).  The data were obtained in the absence of THz radiation applying a current of $i = 10^{-8}\text{A}$ modulated with a frequency of $12 \,\text{Hz}$.  Note that the back gate dependence is asymmetric in respect to the CNP.  For different sample cool-downs the CNP position $U_{\rm CNP}$  slightly shifts. Thus, to compare various measurements we use the effective gate voltage $U^{\rm eff}_{\rm g} = U_{\rm g} - U_{\rm CNP}$. Varying the effective gate voltage from -10 to 10~V the carrier density determined by Hall measurements can be tuned from $p = 5 \times 10^{11}$ to $n = 7.5 \times 10^{11}$ cm$^{-2}$. 
The carrier density changes linearly with $n\,  [\text{cm}^{-2}] = - 0.9 \times 10^{11} U_\mathrm{g}^{\mathrm{eff}}$~[V] and $p\,  [\text{cm}^{-2}] = 0.55 \times 10^{11} U_\mathrm{g}^{\mathrm{eff}}$~[V] and the  mobility $\mu$ at liquid helium temperature is about $10^5$~cm$^2$/Vs.

The photocurrents were excited by applying normally incident linearly polarized radiation of a high power pulsed THz laser. Figure~\ref{fig_1}(d) shows the experimental arrangement. In the measurements we used the radiation of a pulsed NH$_3$ laser~\cite{Ganichev1982,Shalygin2006,Plank2016drag} pumped by a transversely excited atmospheric pressure (TEA) CO$_2$ laser~\cite{Ganichev1982,Ganichev2003}. The  laser provides single pulses of monochromatic radiation with pulse duration in the order of 100~ns, repetition rate of 1~Hz, and peak  powers in the order of hundreds of kW. The peak power of the radiation was monitored with a THz photon-drag detectors~\cite{Ganichev84p20}. The laser operated at frequencies $f = 3.31$, 2.02 or 1.07~THz. Corresponding photon energies are 13.7, 8.4 and 4.4~meV, respectively.  The spot of the terahertz radiation, measured with a pyroelectric camera, has an almost Gaussian profile and is, depending on the used laser line, about 1.5-3~mm in diameter.  Consequently, the micrometer sized Hall bar structures are illuminated homogeneously with radiation intensities up  400~kW/cm$^2$. To vary the orientation of the radiation electric field vector ${\bm E}(t)$ we used crystal quartz $\lambda$/2 wave plates. The direction of the radiation electric field vector in respect to the sample's long side is defined by the azimuth angle $\alpha$, see the inset in Fig.~\ref{fig_2}(b).  To vary the laser radiation intensity, we used two polarizers. The linearly polarized laser radiation first passes through a rotating polarizer resulting in a decrease of the radiation intensity and the rotation of the polarization state. The second polarizer, being at a fixed position, causes a further decrease of the radiation intensity and returns the polarization state to the initial one. By this method, we obtained a controllable variation of the radiation intensity. The measurements have been carried out at temperatures of $T=$ 300 and 4.2\,K. To cool down the samples, they were placed in an optical temperature-regulated continuous flow cryostat with $z-$cut crystal quartz windows. The windows were covered by a black polyethylene film, which is transparent in the THz frequency range, but prevents illumination of the sample by visible or room light. Photocurrents in response to the THz pulses were measured  with a digital oscilloscope as a voltage drop across 50~$\Omega$ load resistors, see Fig.~\ref{fig_1}(d) .

 \begin{figure}
	\centering
	\includegraphics[width=\linewidth]{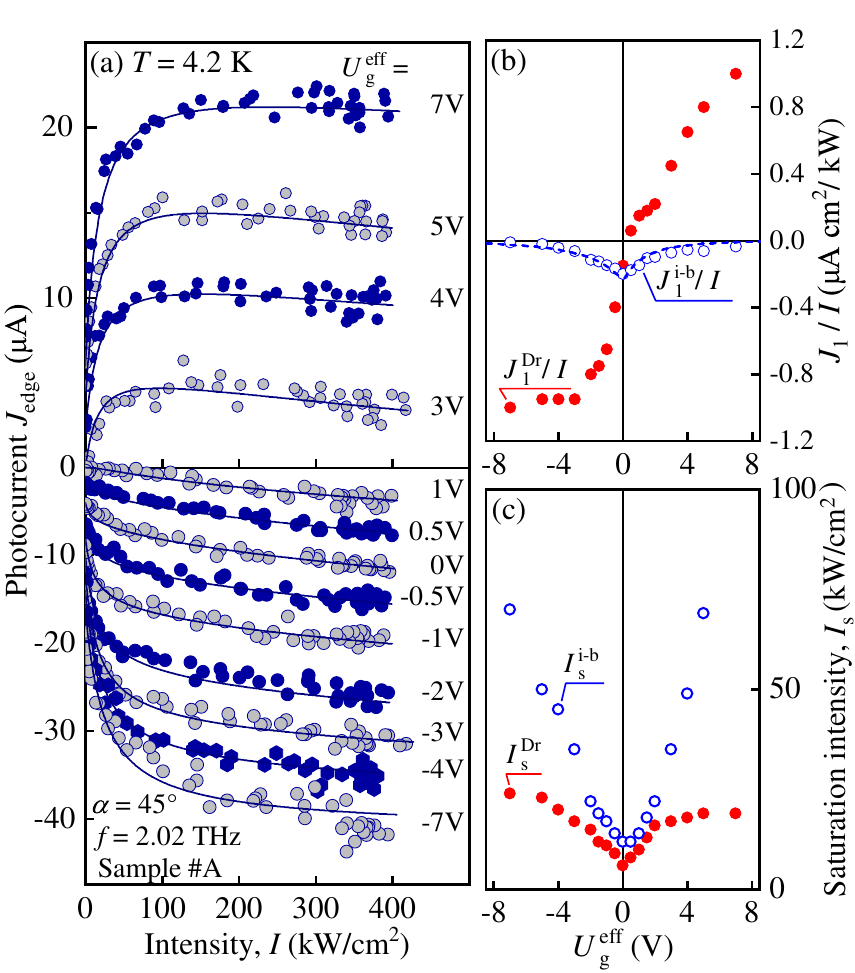}
	\caption{ (a) Intensity dependencies of the edge current excited in sample \#A at liquid helium temperature by radiation with $f = 2.02$~THz. The curves obtained for different values of the effective gate voltage are vertically shifted by 2$\,\upmu$A (except the ones for $U_{\rm g}^{\rm eff}=1$ and 3 V). The solid lines are fits to Eq.~(\ref{both}) with Drude and inter-band amplitudes $J_1^\mathrm{Dr}/I$, $J_1^\mathrm{i-b}/I$,  and saturation intensities $I_s^\mathrm{Dr}$, $I_s^\mathrm{i-b}$ as fit parameters. (b) The magnitudes of the photocurrent contributions caused by Drude ($J^{\rm Dr}_{1}/I$, full circles) and inter-band ($J^{\rm i-b}_{1}/I$, open circles) absorption as a function of the gate voltage. The dashed curve is calculated using Eq.~\eqref{F}. The best fit is obtained assuming an electron temperature $T = 230$~K. Note, that the substantial electron gas heating enables the inter-band absorption at large gate voltages, despite the fact that the final state of such transitions lies below the Fermi energy. (c) Gate dependencies of the saturation intensities of these contributions: $I^{\rm Dr}_s$ (full circles) and $I^{\rm i-b}_s$ (open circles).}
	\label{fig_5}
\end{figure}

\begin{figure}
	\centering
	\includegraphics[]{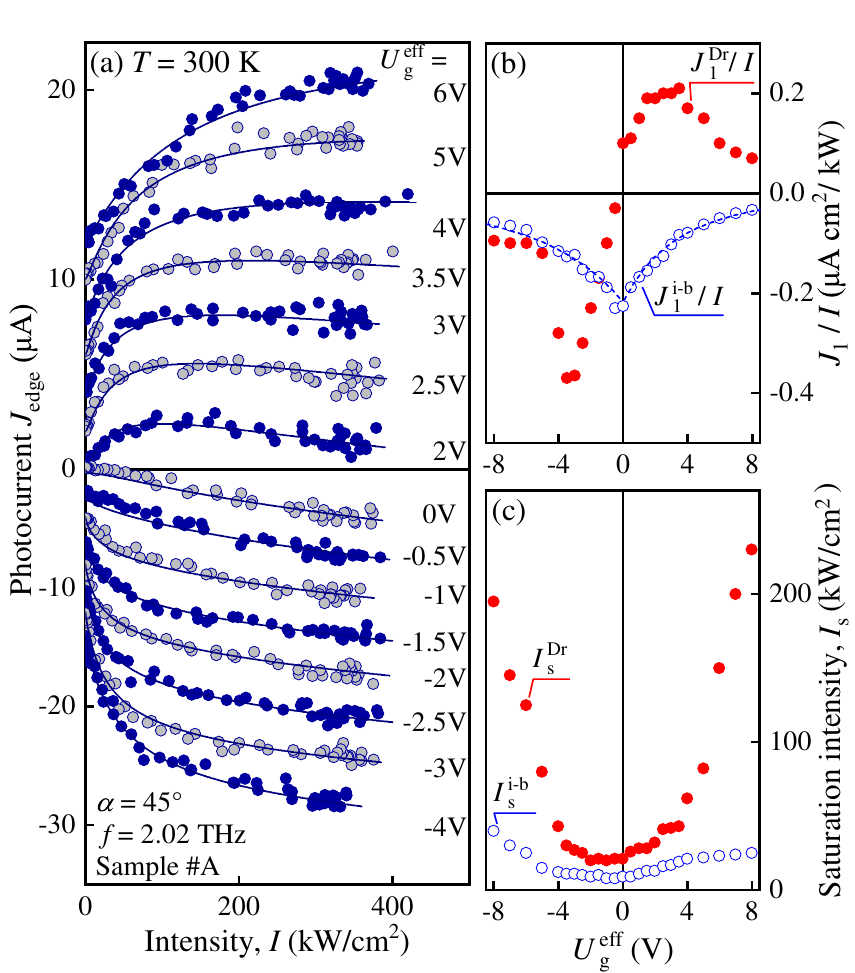}
	\caption{ (a) Intensity dependencies of the edge current excited in sample \#A at room temperature with $f = 2.02$~THz. The curves obtained for different values of the effective gate voltage are vertically shifted by 2$\,\upmu$A (except the ones for $U_{\rm g}^{\rm eff}=1$ and 3 V). Curves are fits using Eq.~(\ref{both}) with Drude and inter-band amplitudes $J_1^\mathrm{Dr}/I$, $J_1^\mathrm{i-b}/I$,  and saturation intensities $I_s^\mathrm{Dr}$, $I_s^\mathrm{i-b}$ as fit parameters.   (b) Photocurrent contributions caused by Drude ($J^{\rm Dr}_{1}/I$, full circles) and inter-band ($J^{\rm i-b}_{1}/I$, open circles) absorption as a function of the gate voltage. The dashed curve is calculated after Eq.~\eqref{F}. The best fit is obtained assuming an electron temperature $T = 370$~K.  (c) Gate dependencies of the saturation intensities of these contributions: $I^{\rm Dr}_s$ (full circles) and $I^{\rm i-b}_s$ (open circles). }
	\label{fig_4}
\end{figure}

\section{Results}

By irradiating the Hall bar structures and measuring the signal from the contact pairs belonging to one of the sample edges we detected the photocurrent, which shows highly nonlinear behaviour upon increase of the radiation intensity.  Figure~\ref{fig_2}(a) exemplary shows the intensity dependence of the photocurrent measured for two pairs of contacts belonging to opposite edges. The data are obtained for a radiation electric field vector oriented at an azimuth angle $\alpha = 45^\circ$. The photocurrent first increases linearly with radiation intensity $I$ and then saturates, so that at high intensities it becomes almost constant or even decreases with rising $I$. Importantly, the photocurrent consistently flows in opposite directions for opposite edges. This indicates that the signal primary stems from the edge and not from the graphene bulk, for which the photocurrent projections on the lines AC and GE would have the same direction. 
The edge contribution can be obtained by subtracting the signals measured at the opposite sides, $J_{\rm edge} = (J_{\rm AC}-J_{\rm GE})/2$. The sign and amplitude of the photocurrent are also defined by the relative orientation between the electric field vector and the corresponding edge. The dependence of the signal on the azimuth angle $\alpha$ can be fitted by~\cite{Plank2019,Candussio2020} 
\begin{equation} \label{eq1}
J_{\rm edge} = J_{\rm L} \sin(2\alpha + \psi) + J_{\rm off}. 
\end{equation}
Here, $J_{\rm L}$ and $J_{\rm off}$ are the amplitudes of the polarization dependent photocurrent  and polarization independent offset, respectively, see Fig.~\ref{fig_2}(b). Note that the phase shift $\psi$ is almost zero.  The photocurrent described by the first term of the right hand side of Eq.~\eqref{eq1} is attributed to the edge photogalvanic current, which is discussed in detail in Refs.~\cite{Karch2011,GlazovGanichev_review,Plank2019,Candussio2020,Durnev_pssb2021,Candussio2021,Durnev2019}. In the next section we briefly address the basic physics of this phenomenon. A small polarization independent offset ($J_{\rm off} < J_{\rm L}$) may be caused by photo-thermoelectrics, see, for example in Ref. \cite{Castilla}. Below we focus on the photogalvanic contribution, easy to extract from the experimental data because its polarization dependence yields a photocurrent of opposite sign for azimuth angles $\alpha = 45^\circ$ and $135^\circ$. 

\begin{figure}
	\centering
	\includegraphics[]{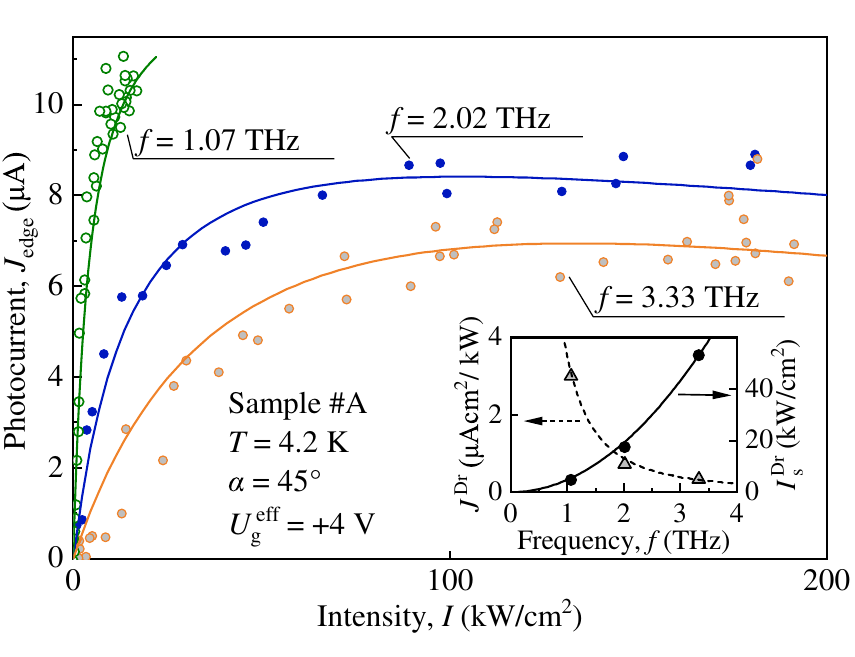}
	\caption{Intensity dependencies of the edge current excited in sample \#A at  liquid helium temperature and high positive gate voltage. The data are presented for three radiation frequencies, $f = 3.33$~ 2.02 and 1.07~THz and $\alpha = 45^\circ$. Curves are fits using Eq.~(\ref{eq2}). Inset shows the frequency dependencies of the magnitude $J^{\rm Dr}$ and saturation intensities of the Drude absorption related photocurrent. Solid and dashed curves show $J^{\rm Dr}\propto 1/\omega^2$ and $I^{\rm Dr}_s \propto \omega^2$, respectively.}
	\label{fig_7}
\end{figure}

\begin{figure}
	\centering
	\includegraphics[]{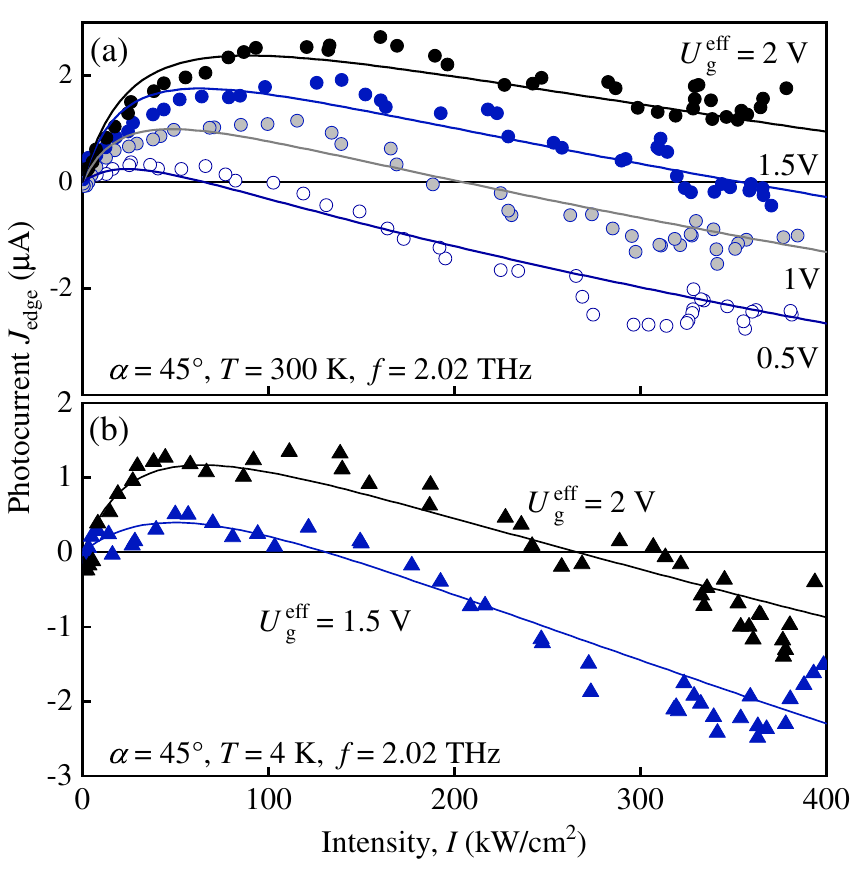}
	\caption{Intensity dependencies of the edge current excited in sample \#A at room (panel a) and liquid helium (panel b) temperatures obtained for small positive gate voltages. The data  are presented for the radiation with $f = 2.02$~THz and $\alpha = 45^\circ$. Solid lines are fits using Eq.~(\ref{eq2}).}
	\label{fig_6}
\end{figure}

The key result of this work is the intensity dependence of the edge photogalvanic current. Our measurements indicate that in a wide range of the back gate voltages the low-power photocurrent depends linearly on light intensity and gradually saturates with increasing intensity, see Figs.~\ref{fig_2}, \ref{fig_5}(a) and \ref{fig_4}(a).  
For the photogalvanic current $J \propto \eta I$ holds, where $\eta$ is the radiation absorbance. Therefore, the observed photocurrent intensity dependence corresponds to a constant absorbance at low power levels (photocurrent scales linearly with the radiation intensity) and a decrease of the absorbance with rising intensity. Figures~\ref{fig_5}(a) and \ref{fig_4}(a) show that the observed nonlinearity changes upon variation of the effective gate voltage and temperature. Furthermore, at large gate voltages we observed that the magnitude of the photocurrent increases with decreasing frequency, and, simultaneously, the photocurrent gets saturated at substantially lower intensities, see Fig.~\ref{fig_7}. The photocurrent signals shown in Figs.~\ref{fig_5}(a) and \ref{fig_4}(a) can be fitted and decomposed into Drude (Dr) and interband (i-b) contributions, using the theory presented in section \ref{nonlinear}. Corresponding plots are shown in Figs.~\ref{fig_5}(b) and \ref{fig_4}(b) for the photocurrent amplitude and in Figs.~\ref{fig_5}(c) and \ref{fig_4}(c) for the saturation intensity. A detailed discussion of these facts will be presented below in Sec.~\ref{discussion}. 

Strikingly, for low positive gate voltages we observed that with increasing radiation intensity the photocurrent reverses its direction, see Fig.~\ref{fig_6}. 
These results reveal that the photocurrent, at least under these conditions, is caused by two competing microscopic mechanisms having different intensity and frequency dependencies. Below we show that these are photogalvanic currents caused by the Drude-like absorption (indirect intraband transitions), $J^{\rm Dr}$, and direct interband transitions between the valence and conduction band, $J^{\rm i-b}$.

\begin{figure}
	\centering
	\includegraphics[]{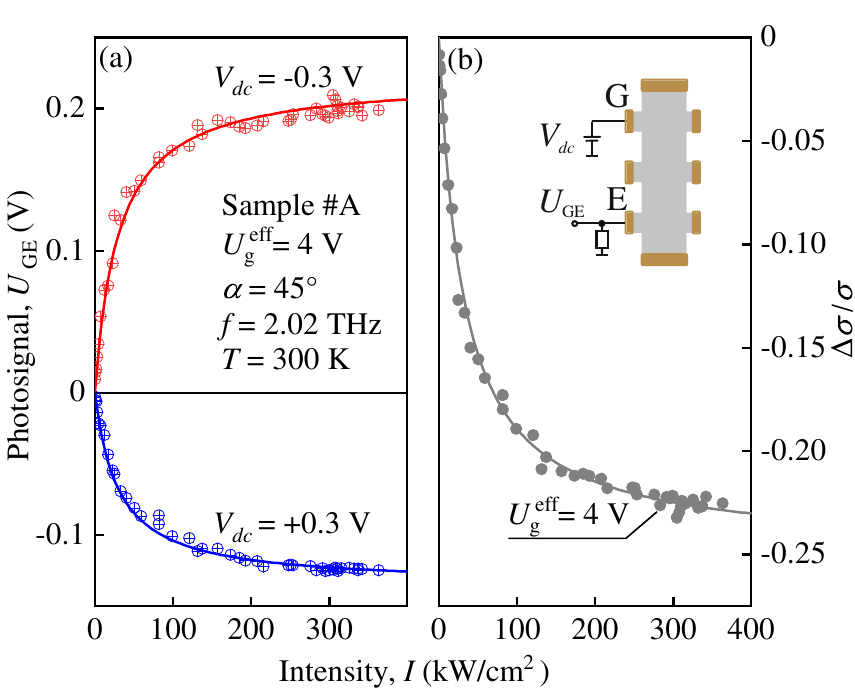}
	\caption{(a) Intensity dependence of the signal measured in the photoconductivity set-up in which a dc bias voltage $V_{dc}$ is applied to the sample. The data are obtained for $f=2.02$~THz, $U_{\rm g}^{\rm eff} = -4$~V and the angle $\alpha =45^\circ$. Fits are after $U_{\rm GE} \propto I/(1 + I/I_s^\mathrm{Dr})$.  
	Panel (b) displays the radiation induced change of the sample conductivity $\Delta \sigma$ normalized by the dark conductivity $\sigma$. The full line is a fit after Eq.~\eqref{eq2}.
	The inset shows the set-up used for the photoconductivity measurements.
}
	\label{fig_8}
\end{figure}

As we will show below, the saturation of the  $J^{\rm Dr}$ contribution is caused by electron gas heating bleaching the radiation absorption. To provide an experimental proof of the strong radiation induced heating of the charge carriers, we additionally studied the photoresponse in the photoconductivity setup. For that we applied a $dc$ bias voltage $V_{dc}$  to the sample, see the inset in Fig.~\ref{fig_8}(b). Figure~\ref{fig_8} shows that for opposite polarities of $V_{dc}$  the photoresponse changes its sign. This proves that under these conditions the radiation induced change of the conductivity  (bolometric effect) dominates the signal. The asymmetry of the signal magnitudes for opposite  $V_{dc}$ polarities, see Fig.~\ref{fig_8}(a),  is caused by the contribution of the photogalvanic curent discussed above. Figure~\ref{fig_8}(b) shows the change of the sample conductivity normalized to the dark conductivity. The conductivity decreases upon irradiation. This behaviour is caused by the so-called negative $\mu-$photoconductivity mechanism for which heating of the charge carriers reduces their mobility~\cite{Ganichevbook}. The decrease of the carrier mobility with increasing electron gas temperature is in agreement with transport results, see e.g.~Ref.~\cite{Sarkar2015} and is due to scattering on acoustic phonons \cite{Wang2013}.
Alike a photocurrent at large gate voltages the photoconductive signal saturates with rising intensity. The data can be well fitted by  
\begin{equation} \label{eq2}
\Delta\sigma/\sigma \propto I/(1 + I/I_s^{\rm Dr})\, ,
\end{equation}
where $I_s^{\rm Dr}$ is the saturation intensity. Note that the  change of sign of the signal with increasing radiation intensity, which is detected in the photocurrent experiments (Fig.~\ref{fig_6}),  is absent in the photoconductivity measurements.

\begin{figure}
	\centering
	\includegraphics[width=\linewidth]{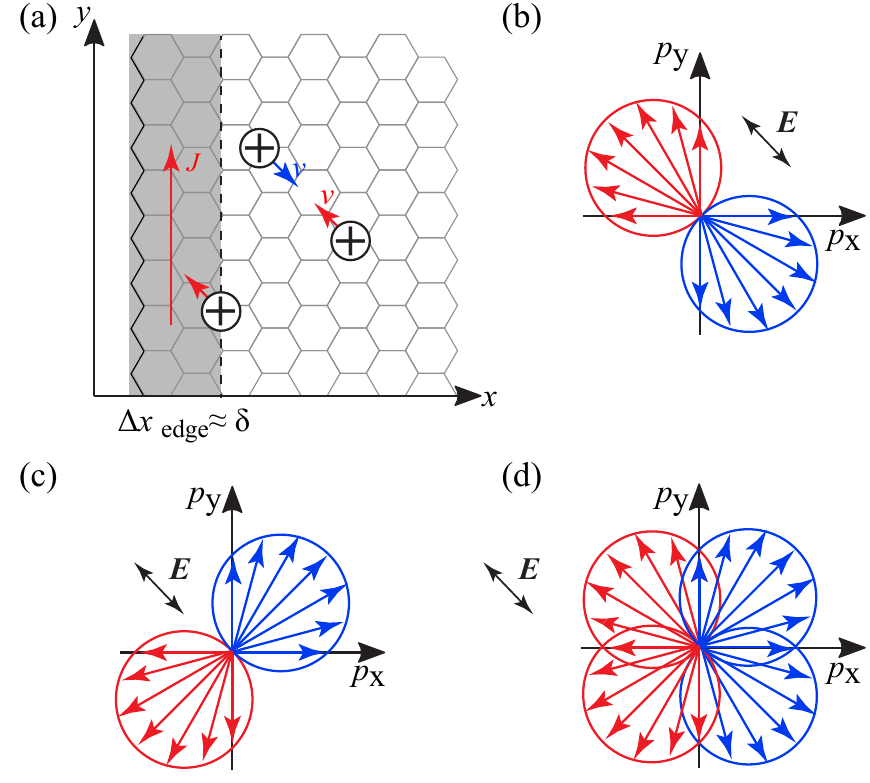}
	\caption{Model of the edge current formation. The current originates  from the alignment of the free carrier momenta by the linearly polarized THz electric field. Blue and red arrows illustrate the anisotropy in the distribution of carrier momenta ${\bm p}$ and, consequently, the velocities ${\bm v}$. (a) The carrier fluxes (red and blue tilted arrows) induced by the optical alignment in stripes of the mean free path width $\delta$ at the sample edge ($\Delta x_{\rm{edge}}$) and in the sample bulk. In the edge vicinity, the uncompensated bottom right flux drives a net electric current $J$ flowing along the edge, illustrated by a vertical red arrow. In the bulk the fluxes compensate each other and a bulk current is absent. After~Ref.~\cite{Candussio2020}. (b-c) Optical  alignment of carrier momenta induced by linearly polarized terahertz radiation resulting in an anisotropy of the carrier movement. The alignment is plotted for Drude absorption (b) and direct interband optical transitions (c). Panel (d) shows the momentum distribution at two-photon interband transitions.} 
	\label{model1}
\end{figure}

\section{Theory}
\label{nonlinear}

At low intensities, the THz radiation induced edge photocurrent caused by Drude-like absorption was studied in mono- and  bi-layer  graphene in Refs.~\cite{Karch2011,GlazovGanichev_review,Plank2019,Durnev2019,Candussio2020,Durnev_pssb2021,Candussio2021,Durnev2021PRB}. It was shown that the $dc$ edge current is induced by the high-frequency electric field of the incident THz radiation due to the broken charge symmetry at the edge. This phenomenon can be viewed as a $dc$ conversion of the bulk $ac$ electric current at the edges. The photocurrent is caused by the  alignment of the free carrier momenta by the high-frequency electric field and subsequent scattering of the carriers at the edge, see Fig.~\ref{model1}. The phenomenon of optical alignment is known for interband and intraband transitions in semiconductors and graphene. The specificity of this process is that during intraband Drude-like optical transitions the quasimomenta get aligned along $\bm E$ ~\cite{Tarasenko2011, Candussio2020, Durnev2021PRB}, whereas for the interband transitions in graphene they are aligned orthogonally to $\bm E$ \cite{Durnev2021PRB,Golub2011,Hartmann2011}, see Fig.~\ref{model1}(b) and (c). The model reveals that the direction of the edge current is defined by the orientation of the $ac$ linearly polarized radiation	electric field with respect to the edge. The dependence of the signal on the azimuth angle $\alpha$ is given by~\cite{Plank2019,Candussio2020} 
	\begin{equation} \label{fit-alpha}
	J = J_{\rm L} \sin 2\alpha \,, 
	\end{equation}
which, as addressed above, describes the experiment, see Fig.~\ref{fig_2}(b).

For Drude absorption, the sign of the edge photocurrent depends -- besides the direction of
the radiation's electric field relative to the sample edge -- on the carrier's
charge $q$ being opposite for $p$- and $n$-type conductivity: $J_y \propto q^3$. The quasi-classical kinetic theory of the low-intensity edge photogalvanic effect in two-dimensional materials with parabolic energy spectrum was developed in~\cite{Candussio2020,Durnev_pssb2021}. Similar calculations of the edge current in graphene with linear energy spectrum yield~\cite{DurnevTarasenko}
\begin{equation} 
\label{eqDr0}
J_1^{\rm Dr} = - \frac{q^3 v_0^2 \tau^3}{2 c n_{\omega} \hbar^2} \frac{3 + (\omega \tau)^2}{[1+(\omega\tau)^2]^2} I \sin 2\alpha \,,
\end{equation}
where  $v_0$ is the velocity in graphene, 
$c$ is the  speed of light, and $n_{\omega}$ is the refractive index of the dielectric medium surrounding graphene. 
 Note that Eq.~(\ref{eqDr0}) is given for the single relaxation time ($\tau$) approximation, for details see Ref.~\cite{Durnev_pssb2021}.

In the case of inter-band absorption for Dirac fermions, optical transitions result in the generation of both types of carriers, electrons and holes, therefore, for systems characterized by electron-hole symmetry, the photocurrent is absent~\cite{Durnev2021PRB}. 
In real graphene flakes the electron-hole symmetry may be lifted due to differences in the scattering time or inhomogeneous doping, causing imbalanced electron and hole photocurrent contributions and enabling photocurrents due to the direct transitions. 
Note, that electron-hole symmetry breaking in charge transport caused by different momentum relaxation times of carriers in the conduction, $\tau_c$, and valence, $\tau_v$, bands was frequently reported in the past, see e.g. Ref.~\cite{Novikov2007,Klatt2010,Obraztsov2014,Li2017}.

While the edge photocurrent has been studied in a regime where it depends
linearly on the light intensity, the nonlinear regime has not been addressed so
far. Here we  develop the theory of edge photogalvanics at interband transitions at arbitrary intensities. We derive the dependence of the edge photocurrent on the light's polarization state and intensity. Afterwards, in Sec.~\ref{discussion}, we discuss the nonlinearity of the edge photogalvanics caused by Drude-like absorption and, combining all these results, analyze the observed complex nonlinear behaviour of the total edge photocurrent.

\subsection{Nonlinear edge photocurrent at interband transitions}
 
The nonequilibrium occupations in the conduction and valence bands $f^{c,v}_{\bm p}$ are found from the kinetic equations, which have the following form in the steady state 
\begin{equation}
\label{kin_eq}
v_x^{c,v} \pdv{f^{c,v}_{\bm p}}{x} +
{f^{c,v}_{\bm p}-\left<f^{c,v}_{\bm p}\right>\over \tau_p^{c,v}} 
+
{f^{c,v}_{\bm p}- f^{c,v}_0 \over \tau_\varepsilon^{c,v}}= \pm G_{\bm p}(f^v_{\bm p}-f^c_{\bm p}).
\end{equation}
Here $x$ is the direction perpendicular to the edge, $\bm v^{c,v}$ is the electron velocity in the conduction or valence band, angular brackets mean averaging over directions of the momentum $\bm p$ at a fixed energy, 
$\tau_p^{c,v}$ and $\tau_\varepsilon^{c,v}$ are the momentum and energy relaxation times of photocarriers in the corresponding band, and
$f_0^{c,v}$ are equilibrium distributions.
The interband generation rate $G_{\bm p}$ is given by
\begin{equation}
\label{G_k}
G_{\bm p} = {2|M_{cv}(\bm p)|^2/\tau \over [\varepsilon_c(p)-\varepsilon_v(p)-\hbar\omega]^2 + (\hbar/\tau)^2}
\end{equation}
with $M_{cv}(\bm p)$ being the matrix element of the direct optical transition, $\varepsilon_{c,v}$ being the conduction- and valence-band dispersions, and 
the relaxation rate is given by a half-sum of the total relaxation rates in the bands:
\begin{equation}
{1\over \tau} = {1\over 2} \qty({1\over \tau_c} + {1\over \tau_v}),
\qquad
{1\over \tau_{c,v}} = {1\over \tau_p^{c,v}} + {1\over \tau_\varepsilon^{c,v}}.
\end{equation}
Below we assume the electron-hole symmetry of the energy spectrum in graphene implying that ${\varepsilon_v(p)=-\varepsilon_c(p)}$ and $\bm v^v = -\bm v^c$. However, we take into account that scattering in the conduction and valence bands
is different.

The nonequilibrium conduction- and valence-band distribution functions are conveniently presented as $f^{c,v}_{\bm p}=f^{c,v}_0 \pm \Delta f_{\bm p}^{c,v}$, where $\Delta f_{\bm p}^{c,v}$ are the radiation-induced corrections. 
The kinetic equations for the corrections have the form
\begin{equation}
\label{kin_eq1}
\pm v_x \pdv{\Delta f_{\bm p}^{c,v} }{x} + {\Delta f_{\bm p}^{c,v}\over \tau_{c,v}} 
= G_{\bm p} \qty[ \mathcal{F}-\qty(\Delta f_{\bm p}^c+\Delta f_{\bm p}^v )] + {\left<\Delta f_{\bm p}^{c,v} \right> \over \tau_p^{c,v}} .
\end{equation}
Here $v_x=v_x^c$ and 
\begin{equation}
\label{F_def}
{\cal F} = f_0\qty(-{\hbar\omega/ 2})-f_0\qty({\hbar\omega/ 2})
\end{equation}
is the difference of occupations of the initial and final states in equilibrium.
The nonlinearity is taken into account by the term $G_{\bm p}\qty(\Delta f_{\bm p}^c+\Delta f_{\bm p}^v )$ on the right-hand side similar to homogeneous systems~\cite{Nonlin_abs_Agarwal,pssb_2019}.

The electric current along the edge is given by
\begin{equation}
\label{J_def}
J = e \sum_{\nu,\bm p} v_y \int\limits_0^\infty \dd x \qty(\Delta f_{\bm p}^c+ \Delta f_{\bm p}^v )_a,
\end{equation}
where $e<0$ is the electron charge, $\nu$ labels spins and valleys,
and the subscripts ``$a$'' and ``$s$'' denote the asymmetric and symmetric in $p_y$ parts of the function:
\[F_{a,s}(\bm p)\equiv [F(p_x,p_y) \mp F(p_x,-p_y)]/2.\]
Analysis shows that the last term in the kinetic Eq.~\eqref{kin_eq1}
does not result in  asymmetric parts of the distributions, 
and we ignore this term below. 

Introducing the asymmetric in $p_y$ functions
\begin{equation}
\Delta f_{\bm p}^{(\pm)}(x) = \qty[\Delta f_{\bm p}^c(x) \pm \Delta f_{\bm p}^v(x)]_a,
\end{equation}
we see that the photocurrent is determined by $\Delta f_{\bm p}^{(+)}$. 
The kinetic equations for $\Delta f_{\bm p}^{(\pm)}$ have the following forms
\begin{align}
&\qty(v_x \partial_x+{1\over \tau_-})\Delta f_{\bm p}^{(-)} + {\Delta f_{\bm p}^{(+)} \over \tau} =   2G_{\bm p,a}\mathcal{F}-2G_{\bm p,s}\Delta f_{\bm p}^{(+)}, \nonumber \\
&\qty(v_x \partial_x+{1\over \tau_-})\Delta f_{\bm p}^{(+)} + {\Delta f_{\bm p}^{(-)} \over \tau} = 0,
\end{align}
where $1/ \tau_- = \qty({1/ \tau_c} - {1/ \tau_v})/2$.
From these two equations we obtain one closed equation for $\Delta f_{\bm p}^{(+)}$:
\begin{equation}
\qty[\qty(v_x \partial_x+{1\over \tau_-})^2 - {1\over \tau^2} - {2G_{\bm p,s}\over \tau}]\Delta f_{\bm p}^{(+)} = 
-{2G_{\bm p,a} \mathcal{F}\over \tau} .
\end{equation}
Its solution has the form
\begin{equation}
\Delta f_{\bm p}^{(+)} = \mathcal{F} {2 G_{\bm p,a}\bar{\tau}\over 1+2 G_{\bm p,s}\bar{\tau}}
+ F_0(\bm p) \exp[-\lambda(\bm p)x].
\end{equation}
Here 
\begin{equation}
\bar{\tau}={\tau_c+\tau_v\over 2},
\end{equation}
the function $F_0(\bm p)$ is determined from the boundary conditions at the edge $x=0$, and $\lambda(\bm p)$ is the positive root of the corresponding characteristic equation:
\begin{equation}
\lambda = {\sqrt{1+2 G_{\bm p,s}\tau}\over \abs{v_x}\tau} + {1\over v_x\tau_-}.
\end{equation}
At low intensity, $G_{\bm p}\bar{\tau} \to 0$, we get 
\begin{equation}
\lambda(\bm p) = {1\over \abs{v_x}} \qty[{\Theta(p_x)\over\tau_c} + {\Theta(-p_x)\over \tau_v}],
\end{equation}
where $\Theta(p)$ is the Heaviside function.
It means that at $p_x>0$ the function $\Delta f^{(+)}_{\bm p}$ is the correction to the distribution function in the conduction band, while at $p_x<0$ it is the correction in the valence band where electrons have positive velocity.

For the diffusive boundary condition $\Delta f^{(+)}_{\bm p}(x=0) = 0$
we obtain
\begin{equation}
\label{d_f}
\Delta f^{(+)}_{\bm p} =  \mathcal{F}  {2 G_{\bm p,a}\bar{\tau}\over 1+2 G_{\bm p,s}\bar{\tau}}\qty[1-\text{e}^{-\lambda(\bm p)x}].
\end{equation}
The first, $x$-independent
term in Eq.~\eqref{d_f} does not contribute to the current because $G_{\bm p}$ is an even function of $\bm p$ in graphene ($G_{\bm p,a}$ being by definition odd in $p_y$ is also odd in $p_x$).
Therefore, we obtain from Eq.~\eqref{J_def} the electric current in the form
\begin{equation}
\label{J1}
J = 2e \mathcal{F}\bar{\tau}^2 \sum_{\nu,\bm p} 
{G_{\bm p,a}v_y \qty(v_x\tau/\tau_- -\sqrt{1+2 G_{\bm p,s}\tau}\abs{v_x})\over \qty(1+2 G_{\bm p,s}\bar{\tau})^2}.
\end{equation}
Analysis shows
that 
the term with a square root
in Eq.~\eqref{J1} yields no contribution to the photocurrent.
As a result we obtain
\begin{equation}
\label{J}
J=e \mathcal{F}{\tau_c^2-\tau_v^2\over 2} \sum_{\nu,\bm p}  {v_y v_xG_{\bm p,a} \over (1+2 G_{\bm p,s}\bar{\tau})^2}
.
\end{equation}

It follows from Eq.~\eqref{J} that at low intensity, when $G_{\bm p,s}\ll 1/(2\bar{\tau})$, the edge current is given by
\begin{equation}
\label{J0}
J_1 
= e \mathcal{F}{\tau_c^2-\tau_v^2\over 2} \sum_{\nu,\bm p}  v_y v_xG_{\bm p}
,
\end{equation}
which coincides with the expression obtained in the linear in the intensity regime in Ref.~\cite{Durnev2021PRB}. Note that, for nondiffusive reflection from the edge, the relaxation times squared in Eq.~\eqref{J0} are changed to $\tau_{c,v}^2\to \tau_{c,v}^2(1+\zeta_{c,v})$, where $\zeta_{c,v}$ are the dimensionless parameters taking into account that the scattering may be partially specular.From Eqs.~\eqref{J} and~\eqref{J0} we obtain that the edge current at any intensity can be expressed as follows
\begin{equation}
\label{J_J0}
J = J_1 {\sum\limits_{\bm p} v_y v_x G_{\bm p}  (1+2 G_{\bm p,s}\bar{\tau})^{-2}
	\over \sum\limits_{\bm p}v_y v_x  G_{\bm p}}.
\end{equation}

Note that the ratios $\tau_\varepsilon^{c,v}/\tau_p^{c,v}$ cancel  from the edge photocurrent which is governed by the total relaxation times in the bands $\tau_{c,v}$. This is in contrast to the photocurrent in infinite homogeneous systems where the relation between the energy and momentum relaxation rates changes strongly the high-intensity behavior of the circular photocurrent even at a fixed total relaxation rate~\cite{pssb_2019}. The difference is caused by the microscopic processes of the photocurrent formation: while the edge photocurrent is formed via edge-scattering of the even electron distribution shown in Fig.~\ref{model1}(c), the circular photocurrent in homogeneous gyrotropic systems is generated in the moment of excitation. Formally, the difference is in the generation rate $G_{\bm p}$ which is even in $\bm p$ in graphene but it is odd in $\bm p$ in, e.g., Weyl semimetals~\cite{pssb_2019}.

\subsection{Edge photocurrent in graphene}

We apply the formalism derived above to the calculation of the edge photocurrent in graphene. Electron states in the conduction and valence bands of graphene
are described by the effective Hamiltonian linear in momentum
\begin{equation}
{\cal H} =  v_0 \bm \sigma \cdot \bm p,
\end{equation}
where $\sigma_{x,y}$ are the Pauli matrices, $x,y$ are directions in the graphene plane, and $v_0$ is the  velocity. 
In this model the energy dispersions are ${\varepsilon_c=-\varepsilon_v= v_0 p}$, and 
the direct optical transition matrix element is given by
\begin{equation}
\label{M_cv}
M_{cv}(\bm p) = {ev_0 \over \omega}{[\bm E \times \bm p]_z \over p}\,,
\end{equation}
where $\bm E$ is the complex amplitude of the radiation electric field. 
Therefore the symmetric and asymmetric in $p_y$ parts of
the generation rate~\eqref{G_k} read as
\begin{align}
\label{G_graphene}
&G_{\bm p,s}\bar{\tau} = {\mathcal{E}^2\over \Delta^2 + 1} \qty(1-\cos{2\varphi_{\bm p}}\cos{2\alpha}),
\\
&G_{\bm p,a}\bar{\tau} = -{\mathcal{E}^2\over \Delta^2 + 1}\sin{2\varphi_{\bm p}}\sin{2\alpha},
\end{align}
where $\Delta=(2v_0p-\hbar\omega)\tau/\hbar$,  
$\alpha$ is the angle between $\bm E$ and the edge, $\varphi_{\bm p}$ is the polar angle of $\bm p$, and
we introduce the dimensionless radiation electric field amplitude defined as
\begin{equation}
{\cal E} = \sqrt{\tau_c\tau_v}{v_0 |e\bm E|\over \hbar \omega}.
\end{equation}
The low-intensity photocurrent is given by
\begin{equation}
\label{J1LG}
J_1 = \sin{2\alpha} {e\mathcal{F}\omega(\tau_c^2-\tau_v^2)\over 16\tau_c\tau_v} \mathcal{E}^2.
\end{equation}
From Eq.~\eqref{J_J0}
at $\omega\tau \gg 1$
we obtain the nonlinear edge photocurrent in graphene in the following form:
\begin{align}
&{J\over J_1} = {2\sqrt{1+4\mathcal{E}^2\cos^2{\alpha}}\over 3\pi \mathcal{E}^4 \cos^2{2\alpha}} 
\nonumber \\ 
&\times \qty[ {1+\mathcal{E}^2 - 2\mathcal{E}^4\sin^2{2\alpha} \over {1+4\mathcal{E}^2\cos^2{\alpha}}} \text{K}(m)  + (\mathcal{E}^2-1)\text{E}(m) ].
\end{align}
Here $\text{E}(m)[\text{K}(m)]=\int_0^{\pi/2}\dd \theta \qty(1-m\sin^2{\theta})^{\pm 1/2}$ are the complete elliptic integrals, and we introduced the notation 
\begin{equation}
m= {4\mathcal{E}^2\cos{2\alpha}\over 1+4\mathcal{E}^2\cos^2{\alpha}}. \nonumber
\end{equation}
At $\alpha=\pi/4$ the expression for the photocurrent simplifies to
\begin{equation}
\label{ibnonlinear}
{J \over J_1 } = {\mathcal{E}^2+1\over (2\mathcal{E}^2+1)^{3/2}}.
\end{equation}
At low and high intensity we have asymptotes:
\begin{align}
\label{asympt}
&{J \over J_1} \approx 1 - 2\mathcal{E}^2 + {9\over 16}(\cos{4\alpha}+9)\mathcal{E}^4  \quad (\mathcal{E}\to 0),
\\
&{J \over J_1}\approx {4 \abs{\cos{\alpha}} \qty[\text{E}\qty({\cos{2\alpha}\over\cos^2{\alpha}})-2\sin^2{\alpha} \text{K}\qty({\cos{2\alpha}\over\cos^2{\alpha}})]\over 3\pi\mathcal{E} \cos^2{2\alpha}} \quad (\mathcal{E}\to \infty). \nonumber
\end{align}
We see that the edge photocurrent ${J \propto \sqrt{I}}$ at high intensity. Interestingly, the high-intensity expression is frequency-independent (at $\mathcal{F}=1$) similar to the circular photocurrent in homogeneous gyrotropic systems~\cite{Matsyshyn2021}.

\begin{figure}[h]
	\centering
			\includegraphics[width=\linewidth]{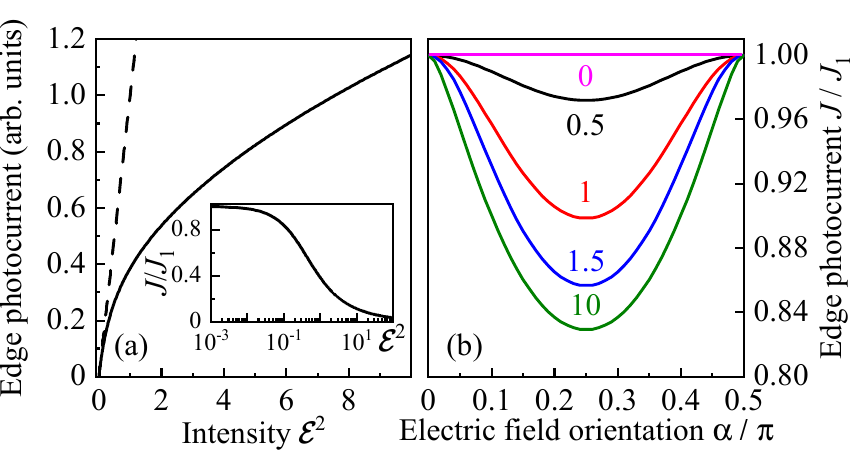}
	\caption{ (a) The dependence of the edge photocurrent on the light intensity at $\alpha=\pi/4$. The dashed line is the low-intensity current $J_1$. Inset shows the ratio  $J/J_1$ as a function of $\mathcal{E}^2$ on a semilogarithmic scale. 
		(b) Deviation of the edge current from the $\sin{2\alpha}$ dependence given by the ratio  $J/J_1$ at different values of the electric field amplitude ${\mathcal{E}}$ indicated in the figure. The ratios $J/J_1$ are normalized to their values at $\alpha=0$.
	}
	\label{fig_Theor}
\end{figure}

The dependence of the edge photocurrent on the light intensity is shown in Fig.~\ref{fig_Theor}(a).
From that it is clear, that the linear in intensity regime where $J\approx J_1 \propto I$ is realized at $\mathcal{E}^2 \leq 0.2$ only.
The dependence of the ratio $J/J_1$ on the polarization direction is shown in Fig.~\ref{fig_Theor}(b). It demonstrates that the dependence $J(\alpha)$ at high intensity differs from $J_1(\alpha)\propto \sin{2\alpha}$ but this deviation does not exceed 15~\%.
This is also seen from  Eq.~\eqref{asympt} where 
the dependence of the ratio $J/J_1$ on the polarization state appears in the fourth order in $\mathcal{E}$ only, and its amplitude is small in comparison to the $\alpha$-independent part.

\subsection{Two-photon absorption in graphene}

So far we considered one-photon processes. However, in the THz frequency range, the two-photon absorption can be also strong and comparable with the one-photon absorption, see Refs.~\cite{Ganichevbook,Ganichev1983,Ganichev1993,Yang2011,Rioux2011}. 

To account for this effect, we consider two-photon direct interband optical transitions $v \to c$.
The corresponding matrix element
is given by
\begin{equation}
M_{cv}^{(2)} = M_{cv} \qty({V_{vv}\over \hbar \omega} + {V_{cc}\over \hbar \omega + \varepsilon_v-\varepsilon_c}).
\end{equation}
Here $M_{cv}$ is the one-photon direct optical transition matrix element, and the intraband matrix elements of the electron-photon interaction are $V_{nn}= (ie/\omega)\bm E \cdot \bm v^n$ ($n=c,v$).
Using the energy conservation law for two-photon absorption $\varepsilon_c-\varepsilon_v=2\hbar\omega$, we obtain
\begin{equation}
M_{cv}^{(2)} = -M_{cv}{ie\over \hbar \omega^2} \bm E \cdot (\bm v^c-\bm v^v).
\end{equation}

For graphene we have $\bm v^c = -\bm v^v = v_0{\bm p/p}$, and $M_{cv}$ is given by Eq.~\eqref{M_cv}. Therefore we get
\begin{equation}
\label{M_2_square}
\qty|M_{cv}^{(2)}|^2 =   {(ev_0)^4 \over \hbar^2 \omega^6} |\bm E|^4 \qty[1-P_L^2\cos^2{2(\varphi_{\bm p}-\alpha)}],
\end{equation}
where $P_L$ is the linear polarization degree.
%
The 
two-photon absorbance $\eta^{(2)}$ is found from the Fermi golden rule
\begin{equation}
{\eta^{(2)} I \over 2\hbar \omega}= {2\pi \over \hbar} \mathcal{F}_2\sum_{\nu, \bm p} \qty|M_{cv}^{(2)}(\bm p)|^2 \delta(2 v_0 p - 2\hbar \omega),
\end{equation}
where ${\cal F}_2 = f_0\qty(-{\hbar\omega})-f_0\qty({\hbar\omega})$.
Calculation of the sum yields
\begin{equation}
\label{eta_2_0}
\eta^{(2)} = I \qty({e^2\over \hbar c})^2 {4\pi^2 v_0^2 \over \hbar \omega^4} \mathcal{F}_2(1-P_L^2/2) .
\end{equation}
We see that the linear-circular dichroism of the two-photon absorption  takes place  in graphene: 
${\eta_{\text{circ}}^{(2)}/ \eta_{\text{lin}}^{(2)} = 2}$~\cite{Rioux2011}.

The total absorbance of circularly-polarized light  accounting for both one- and two-photon processes
is given by
\begin{equation}
\label{eta_I}
\eta_\text{circ} = {\pi e^2\over \hbar c}\qty({\cal F} + {\cal F}_2 {I\over I_0}),
\end{equation}
where 
the characteristic intensity $I_0$ is defined as
\begin{equation}
I_0 
= {c\over 16\pi}\qty({\hbar\omega^2\over ev_0})^2.
\end{equation}
This expression demonstrates that for an intensity $I \sim I_0$ the two photon absorption is comparable with the one-photon absorption. 
Substituting the carrier velocity  in graphene $v_0 = 10^{8}$~cm/s we obtain
\begin{equation}
I_0 = 1.76 \qty({\hbar\omega\over 10 \: \text{meV}})^4 ~{[\text{kW}/\text{cm}^2]}.
\end{equation}
This yields $I_0=2.3$, 29.3 and 216.4~kW/cm$^2$ for the frequencies $f=1.07$, 2.02 and 3.33~THz, respectively. Consequently, we obtained that, at intensities and frequencies used in this work, the two-photon and one-photon absorbance are comparable.

\begin{figure}
	\centering
	\includegraphics[width=\linewidth]{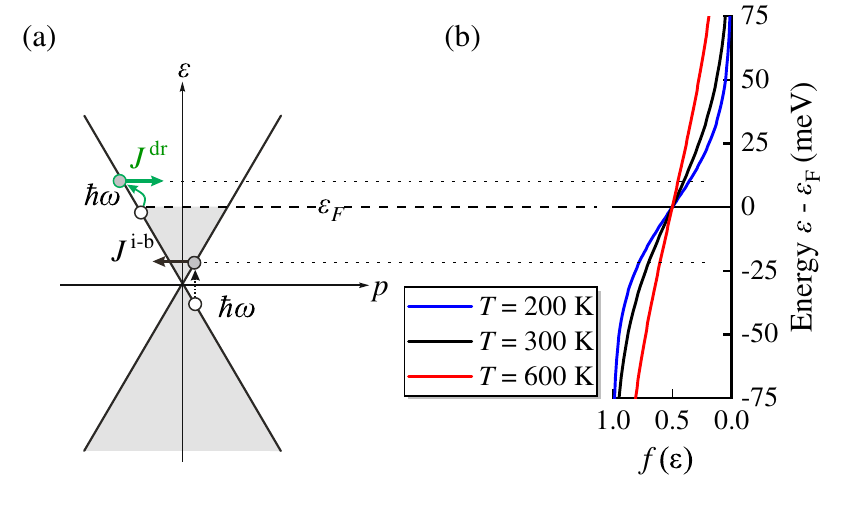}
	\caption{Model illustrating the edge photocurrent formation for Drude optical transitions (bent arrow) and direct inter-band transitions (upward arrow). Open and closed circles illustrate carriers in the initial and final state of the optical transitions, respectively. Right panel illustrates  modification of the Fermi-Dirac distribution due to the electron gas heating. The carrier redistribution is illustrated for the Fermi level lying in the conduction band and three different temperatures.	Despite the final state of the indicated direct transition is below the Fermi level, a substantial heating resulting in a high temperature leads to a depopulation of this state making allowance for the inter-band transition.}
	\label{fig_10}
\end{figure}

However, our analysis shows that the two-photon processes do not yield an edge photocurrent. Indeed, it follows from Eq.~\eqref{M_2_square} that the carrier distribution in the momentum space for linear polarization is described by the fourth Fourier-harmonics, see Fig.~\ref{model1}(d). This means that two-photon processes do not result in an alignment of carrier momenta, in sharp contrast to the one-photon absorption. Therefore, scattering from the edge does not result in the electric current in the case of two-photon absorption.

\section{Discussion}
\label{discussion}

As addressed above, the edge photocurrent may be caused by two microscopic mechanisms involving indirect  intraband optical transitions (Drude like) and direct optical transitions between the valence and  conduction bands. Figure~\ref{fig_10} schematically shows the generation of these photocurrent contributions, panel (a), together with the Fermi-Dirac distribution function plotted for different temperatures, panel (b).
The photocurrent caused by direct optical transitions is proportional to the difference of initial and final states occupancy of the direct transition Eq.~\eqref{F_def}, which is given by 
\begin{equation}
\label{F}
{\cal F} = 
f_0\qty(-{\hbar\omega\over 2})-f_0\qty({\hbar\omega\over 2}) = 
{\sinh({\hbar\omega\over 2k_\text{B}T})\over \cosh({\hbar\omega\over 2k_\text{B}T}) + \cosh({\varepsilon_\text{F}\over k_\text{B}T})}.
\end{equation}
The factor $\mathcal{F}$ depends on the gate voltage via the position of the Fermi level $|\varepsilon_\text{F}| =\hbar v_0 \sqrt{\pi n^*}$, where $n^*$ is $n$ or $p$ for positive or negative effective gate voltages $U_g^{\rm eff}$, respectively.
At low temperature and high effective gate voltages, i.e. high Fermi level position, $\cal F$ becomes small and $J^{\rm i-b}$ is negligible as compared to the photocurrent stemming from the Drude absorption $J^{\rm Dr}$.
In our experiments this corresponds to the traces obtained for liquid helium temperature and gate voltages $\pm 7$~V, see Fig.~\ref{fig_5}(a). Under these conditions, the photocurrent saturates when the radiation intensity is increased. The observed nonlinearity is attributed to electron gas heating resulting in absorption bleaching and, consequently, photocurrent saturation. The bleaching of the Drude-like radiation absorption in graphene has recently been observed and studied in details  applying nonlinear ultrafast THz spectroscopy, reported in Ref. \cite{Mics2015}, where the THz conductivity was investigated by analysing the THz pulse transmission. The range of frequencies (0.4–1.2 THz) and radiation electric fields  (2 – 100 kV/cm)  used in Ref.~\cite{Mics2015} are similar to that used in our work. 
It has been demonstrated that for such THz radiation intensities the electron temperature may increase to several thousand Kelvin,
which results in a decrease of the THz conductivity when increasing the intensity. Within a thermodynamic approach, the results are well described by the statistically determined thermal balance maintained within the entire electron population of graphene. Complementary measurements of the THz photoconductivity, see Fig.~\ref{fig_8}, which for Drude absorption stems from electron gas heating and the associated mobility reduction, confirms that in our experiments the radiation induces a strong electron gas heating. For the data of Ref.~\cite{Mics2015} we find that they are well described by an empirical analytical formula $\eta \propto 1/(1+ I/I_s^{\rm Dr})$, with $I_s^{\rm Dr}$ the saturation intensity being proportional to the reciprocal energy relaxation time and the Drude absorption cross-section.  As the photogalvanic current is proportional to the radiation absorbance  we further use this formula to fit the Drude contribution to the experimental traces as (hereafter $\alpha=45^\circ$)
\begin{equation} \label{eqDr}
J_L^{\rm Dr} = \frac{J_1^{\rm Dr}}{1+ I/I_s^{\rm Dr}}.
\end{equation}

A decrease of the Fermi level position together with electron gas heating enhances the direct inter-band optical transitions and causes the corresponding photocurrent, $J_L^{\rm i-b}$. The total photocurrent is the sum of the two contributions: $J_L^{\rm i-b}$  and $J_L^{\rm Dr}$, see Fig.~\ref{fig_10}. Combining Eqs.~(\ref{ibnonlinear}) and (\ref{eqDr}) we obtain for the intensity dependence of the total photocurrent amplitude, introduced in Eq.~\eqref{fit-alpha}
\begin{equation} \label{both}
J_L = \frac{J_1^{\rm Dr}}{1+ I/I_s^{\rm Dr}}  +  J_1^{\rm i-b} {I/I_s^{\rm i-b}+1\over (2I/I_s^{\rm i-b}+1)^{3/2}},
\end{equation}
where according to Eq.~\eqref{J1LG}
\begin{equation}
\label{J0interband}
J_1^{\rm i-b} = 
{\pi e  \mathcal{F}\omega\over 8c n_\omega} (\tau_c^2-\tau_v^2) \qty({e v_0\over \hbar\omega})^2 I\,,
\end{equation}
%
and the square of the dimensionless electric field is presented in the form
\begin{equation} 
\mathcal{E}^2={I\over I_\mathrm{s}^\mathrm{i-b}}
\end{equation}
with the saturation intensity 
\begin{equation}
\label{Isib}
I_s^{\rm i-b} = {c\over 2\pi \tau_c\tau_v}\qty({\hbar\omega\over e v_0})^2.
\end{equation}
 
Equation~(\ref{both}) describes well all intensity dependencies obtained for different gate voltages, temperatures and radiation frequencies, see Figs.~\ref{fig_5}(a), \ref{fig_4}(a), \ref{fig_7}, and \ref{fig_6}. The effective back gate voltage dependence of the fitting parameters $J_1^{\rm Dr}$, $J_1^{\rm i-b}$, $I_s^{\rm Dr}$, and $I_s^{\rm i-b}$, extracted from the intensity dependencies measured in sample \#A for frequency $f =2.02$~THz, are shown in Fig.~\ref{fig_5} (liquid helium temperature) and Fig.~\ref{fig_4} (room temperature). 

In the following we will discuss the low power amplitudes of the Drude and inter-band photocurrents $J^\mathrm{Dr}_1$ and $J_1^\mathrm{i-b}$, respectively. Figures~\ref{fig_5}(b) and~\ref{fig_4}(b) show that at low temperatures and high gate voltages the Drude photocurrent is larger than the interband photocurrent, see Fig.~\ref{fig_5}(b). Furthermore, while the Drude photocurrent changes its sign at the CNP, the sign of the interband photocurrent is negative in the whole range of the effective back gate voltages  $U_g^{\rm eff}$.  These features are in agreement with Eqs.~(\ref{eqDr0}) and~(\ref{J0interband}): the Drude photocurrent is odd in the carrier charge ($J_1^{\rm Dr}\propto q^3$), whereas the sign of the inter-band photocurrent is caused by the generation of electron-hole pairs, defined only by the difference of the electron and hole momentum relaxation times, see Eq.~(\ref{J0interband}). As addressed in Sec.~\ref{nonlinear}, the observation of  inter-band photogalvanic current reveals that the times  $\tau_c$ and $\tau_v$ are different. Our results for both, liquid helium and room temperatures, demonstrate that for positive gate voltages the inter-band and Drude photocurrent contributions consistently have opposite polarities. Taking into account that for fixed azimuthal angle $\alpha$ the directions of the momentum alignments for these two processes differ by 90$^\circ$, see Fig.~\ref{model1}(b) and (c), we obtain that in our samples $\tau_c > \tau_v$ holds and the inter-band photocurrent is dominated by the photoexcited electrons.  

Equation (\ref{J0interband}) yields that $J_1^{\rm i-b}$ is proportional to the difference between the occupancy of initial and final states of the direct transition $\cal F$, see Eq.~\eqref{F}. Indeed, the fits using $J_1^{\rm i-b}\propto \cal F$, presented in Figs.~\ref{fig_5}(b) and ~\ref{fig_4}(b), describe the gate dependence of this contribution well. To fit these data we took electron gas heating into account, which, despite the small photon energies used in our experiments and high Fermi level position at large effective gate voltages, makes the direct inter-band transition efficient, see Fig.~\ref{fig_10}(b). 

The gate voltage dependence of the Drude photocurrent`s low power amplitude $J_1^{\rm Dr}$ is shown in Figs.~\ref{fig_5}(b) and ~\ref{fig_4}(b). Its overall behaviour is well described by Eq.~(\ref{eqDr0}). Close to the CNP the net electric current vanishes since the electron and hole contributions compensate each other. With the shift of the Fermi level from the Dirac point, the electron gas becomes degenerate and the electric current reaches the value given by Eq.~(\ref{eqDr0}). Further increase of the Fermi energy will result in decrease of the current magnitude if the relaxation times get shorter, see, e.g., Refs.~\cite{Wang2013,Dean2010,Brown2016,Gosling2021,Hirai2014,Banszerus2014}. Figures ~\ref{fig_5}(b) and ~\ref{fig_4}(b)  show that in the vicinity of  the CNP, at which $J_1^{\rm Dr}$ reduces and changes its sign, the photogalvanic current due to direct inter-band transitions dominates the total current.

Now we discuss the saturation intensities of the Drude and inter-band photocurrents $I_s^\mathrm{Dr}$ and $I_s^\mathrm{i-b}$, see Figs.~\ref{fig_5}(c) and ~\ref{fig_4}(c) for low and room temperature, respectively. First of all, as a common feature we observed that the saturation intensities for both mechanisms increase with increase of the effective gate voltage and temperature. Because the saturation intensities are primarily inversely proportional to the reciprocal relaxation times, Eq.~\eqref{Isib}, this observation demonstrates that the times substantially decrease with increase of the gate voltage and temperature. This agrees with the results of Ref.~\cite{Wang2013,Hirai2014,Banszerus2014,Gosling2021} showing that the carrier mobility may reduce with the gate voltage increase.  
Importantly, Eq.~(\ref{both}) shows that the Drude and inter-band contributions are characterized by different intensity dependencies. Indeed for $I \gg I_s^{\rm Dr}$ the corresponding photocurrent contribution becomes independent of the radiation intensity, whereas for $I \gg I_s^{\rm i-b}$  the inter-band photocurrent increases as $J^{\rm i-b} \propto \sqrt{I}$. This difference results in the observed inversion of the photocurrent sign at small positive gate voltages, see Fig.~\ref{fig_6}. At low power the Drude photocurrent is somewhat larger than the inter-band one. Thus the total photocurrent  has a positive sign, see Figs.~\ref{fig_5}(b) and \ref{fig_4}(b). An increase of the radiation intensity saturates the Drude photocurrent, whereas the inter-band photogalvanic current still grows.  Consequently, the total photocurrent becomes dominated by the inter-band contribution and reverses its sign, see Fig.~\ref{fig_6}. The discussed mechanism explains well the observed shift of the inversion intensity with increasing the effective gate voltage, see Fig.~\ref{fig_6}, and its disappearance for high gate voltages,  see Figs.~\ref{fig_5} and~\ref{fig_4}. Indeed, the higher the applied gate voltage, the higher is the relative contribution of the Drude photocurrent and its saturation intensity, see Figs.~\ref{fig_5} and~\ref{fig_4}. Both factors prevent the transition from the Drude contribution to the inter-band one, so that the sign inversion requires higher and higher intensities. 

The data discussed above are obtained applying radiation with frequency $f=2.02$~THz. Intensity dependencies measured for different radiation frequencies are presented in Fig.~\ref{fig_7} (high positive gate voltages). Figure~\ref{fig_7} shows the data for $U_g^{\rm eff} = 4$~V at which the photocurrent is dominated by the Drude absorption, see Fig.~\ref{fig_5}(b). Under these conditions we observed that (i) at low intensities the amplitude of the photocurrent strongly increases upon decrease of the radiation frequency and (ii) for lower frequencies the photocurrent saturates at substantially lower intensities. The inset in  Fig.~\ref{fig_7} shows the frequency dependence of the photocurrent amplitude $J_1^{\rm Dr}$ together with the fit after Eq.~\eqref{eqDr0}, which describes well the observed behaviour. Note that for our experimental conditions $\omega \tau > 1$ and $J_1^{\rm Dr} \propto \omega^{-2}$ holds. The inset in Fig.~\ref{fig_7} also demonstrates that the reduction of frequency results in a decreasing saturation intensity proportional to $\omega^2$. As addressed above, the saturation intensity is proportional to the reciprocal value of the cross-section of the radiation absorbance, which for Drude-like absorption and $\omega \tau > 1$  is proportional to $\omega^{-2}$. 

Finally we compare the experimental saturation intensities $I_s^{\rm i-b}$ of the inter-band photogalvanic current with the theoretical estimation after Eq.~(\ref{Isib}). This expression yields for 
$f=2.02$~THz
\begin{equation}
I_s^{\rm i-b} = {37.2 \over \tau_c\tau_v/\text{ps}^2} ~{[\text{kW}/\text{cm}^2]}.
\end{equation}
For the relaxation times of about 1 ps, which correspond to that obtained for the carrier mobility, we obtain saturation intensities of the order of 50 kW/cm$^2$, which is in excellent agreement with that detected in experiment, see Figs.~\ref{fig_5} and~\ref{fig_4}.

\section{Conclusions}
Our experimental data and theory show that the edge photogalvanic current excited by intense terahertz radiation is characterized by a complex nonlinear intensity dependence. The total photocurrent is shown to be caused by two competing mechanisms related to the Drude-like absorption and direct inter-band optical transitions. Our analysis reveals that the two-photon absorption does not result in photocurrent generation. We demonstrate that at high intensities both kinds of photocurrents saturate with increasing radiation intensities, but are described by different functional behaviour. This causes a complex intensity dependence of the total current, which, depending on gate voltage, temperature and radiation frequency, can either saturate or reverse its sign upon rising intensity. The developed theory of the nonlinear interband photocurrent is in good quantitative agreement with the experimental findings. The obtained characteristics provide an important information for the development of graphene based saturable absorbers for the THz spectral range.

\section{Acknowledgments}
We thank S.A. Tarasenko, M.V. Durnev, and H. Plank for helpful discussions. The support from the Deutsche Forschungsgemeinschaft (DFG, German Research Foundation) – Project-ID 314695032 – SFB 1277  (project A04 and A09), the Elite Network of Bavaria (K-NW-2013-247), the IRAP program of the Foundation for Polish Science (grant MAB/2018/9, project CENTERA), and the Volkswagen Stiftung Program is gratefully acknowledged. L.E.G. thanks the financial support of the Russian Science Foundation (Project 20-12-00147) and the Foundation for the Advancement of Theoretical Physics and Mathematics ``BASIS''. 
K.W. and T.T. acknowledge support from the Elemental Strategy Initiative
conducted by the MEXT, Japan (Grant Number JPMXP0112101001) and  JSPS
KAKENHI (Grant Numbers 19H05790 and JP20H00354).
%

\bibliography{references}

\begin{thebibliography}{67}%
\makeatletter
\providecommand \@ifxundefined [1]{%
 \@ifx{#1\undefined}
}%
\providecommand \@ifnum [1]{%
 \ifnum #1\expandafter \@firstoftwo
 \else \expandafter \@secondoftwo
 \fi
}%
\providecommand \@ifx [1]{%
 \ifx #1\expandafter \@firstoftwo
 \else \expandafter \@secondoftwo
 \fi
}%
\providecommand \natexlab [1]{#1}%
\providecommand \enquote  [1]{``#1''}%
\providecommand \bibnamefont  [1]{#1}%
\providecommand \bibfnamefont [1]{#1}%
\providecommand \citenamefont [1]{#1}%
\providecommand \href@noop [0]{\@secondoftwo}%
\providecommand \href [0]{\begingroup \@sanitize@url \@href}%
\providecommand \@href[1]{\@@startlink{#1}\@@href}%
\providecommand \@@href[1]{\endgroup#1\@@endlink}%
\providecommand \@sanitize@url [0]{\catcode `\\12\catcode `\$12\catcode
  `\&12\catcode `\#12\catcode `\^12\catcode `\_12\catcode `\%12\relax}%
\providecommand \@@startlink[1]{}%
\providecommand \@@endlink[0]{}%
\providecommand \url  [0]{\begingroup\@sanitize@url \@url }%
\providecommand \@url [1]{\endgroup\@href {#1}{\urlprefix }}%
\providecommand \urlprefix  [0]{URL }%
\providecommand \Eprint [0]{\href }%
\providecommand \doibase [0]{https://doi.org/}%
\providecommand \selectlanguage [0]{\@gobble}%
\providecommand \bibinfo  [0]{\@secondoftwo}%
\providecommand \bibfield  [0]{\@secondoftwo}%
\providecommand \translation [1]{[#1]}%
\providecommand \BibitemOpen [0]{}%
\providecommand \bibitemStop [0]{}%
\providecommand \bibitemNoStop [0]{.\EOS\space}%
\providecommand \EOS [0]{\spacefactor3000\relax}%
\providecommand \BibitemShut  [1]{\csname bibitem#1\endcsname}%
\let\auto@bib@innerbib\@empty
\bibitem [{\citenamefont {Novoselov}\ \emph {et~al.}(2006)\citenamefont
  {Novoselov}, \citenamefont {McCann}, \citenamefont {Morozov}, \citenamefont
  {Fal'ko}, \citenamefont {Katsnelson}, \citenamefont {Zeitler}, \citenamefont
  {Jiang}, \citenamefont {Schedin},\ and\ \citenamefont
  {Geim}}]{Novoselov2006}%
  \BibitemOpen
  \bibfield  {author} {\bibinfo {author} {\bibfnamefont {K.~S.}\ \bibnamefont
  {Novoselov}}, \bibinfo {author} {\bibfnamefont {E.}~\bibnamefont {McCann}},
  \bibinfo {author} {\bibfnamefont {S.~V.}\ \bibnamefont {Morozov}}, \bibinfo
  {author} {\bibfnamefont {V.~I.}\ \bibnamefont {Fal'ko}}, \bibinfo {author}
  {\bibfnamefont {M.~I.}\ \bibnamefont {Katsnelson}}, \bibinfo {author}
  {\bibfnamefont {U.}~\bibnamefont {Zeitler}}, \bibinfo {author} {\bibfnamefont
  {D.}~\bibnamefont {Jiang}}, \bibinfo {author} {\bibfnamefont
  {F.}~\bibnamefont {Schedin}},\ and\ \bibinfo {author} {\bibfnamefont {A.~K.}\
  \bibnamefont {Geim}},\ }\bibfield  {title} {\bibinfo {title} {Unconventional
  quantum hall effect and berry's phase of 2$\uppi$ in bilayer graphene},\
  }\href {https://doi.org/10.1038/nphys245} {\bibfield  {journal} {\bibinfo
  {journal} {Nat. Phys.}\ }\textbf {\bibinfo {volume} {2}},\ \bibinfo {pages}
  {177} (\bibinfo {year} {2006})}\BibitemShut {NoStop}%
\bibitem [{\citenamefont {Neto}\ \emph {et~al.}(2009)\citenamefont {Neto},
  \citenamefont {Guinea}, \citenamefont {Peres}, \citenamefont {Novoselov},\
  and\ \citenamefont {Geim}}]{Neto2009}%
  \BibitemOpen
  \bibfield  {author} {\bibinfo {author} {\bibfnamefont {A.~H.~C.}\
  \bibnamefont {Neto}}, \bibinfo {author} {\bibfnamefont {F.}~\bibnamefont
  {Guinea}}, \bibinfo {author} {\bibfnamefont {N.~M.~R.}\ \bibnamefont
  {Peres}}, \bibinfo {author} {\bibfnamefont {K.~S.}\ \bibnamefont
  {Novoselov}},\ and\ \bibinfo {author} {\bibfnamefont {A.~K.}\ \bibnamefont
  {Geim}},\ }\bibfield  {title} {\bibinfo {title} {The electronic properties of
  graphene},\ }\href {https://doi.org/10.1103/revmodphys.81.109} {\bibfield
  {journal} {\bibinfo  {journal} {Rev. Mod. Phys.}\ }\textbf {\bibinfo {volume}
  {81}},\ \bibinfo {pages} {109} (\bibinfo {year} {2009})}\BibitemShut
  {NoStop}%
\bibitem [{\citenamefont {Murakami}\ \emph {et~al.}(2004)\citenamefont
  {Murakami}, \citenamefont {Nagaosa},\ and\ \citenamefont
  {Zhang}}]{Murakami2004}%
  \BibitemOpen
  \bibfield  {author} {\bibinfo {author} {\bibfnamefont {S.}~\bibnamefont
  {Murakami}}, \bibinfo {author} {\bibfnamefont {N.}~\bibnamefont {Nagaosa}},\
  and\ \bibinfo {author} {\bibfnamefont {S.-C.}\ \bibnamefont {Zhang}},\
  }\bibfield  {title} {\bibinfo {title} {Spin-hall insulator},\ }\href
  {https://doi.org/10.1103/physrevlett.93.156804} {\bibfield  {journal}
  {\bibinfo  {journal} {Phys. Rev. Lett.}\ }\textbf {\bibinfo {volume} {93}},\
  \bibinfo {pages} {156804} (\bibinfo {year} {2004})}\BibitemShut {NoStop}%
\bibitem [{\citenamefont {Kane}\ and\ \citenamefont {Mele}(2005)}]{Kane2005}%
  \BibitemOpen
  \bibfield  {author} {\bibinfo {author} {\bibfnamefont {C.~L.}\ \bibnamefont
  {Kane}}\ and\ \bibinfo {author} {\bibfnamefont {E.~J.}\ \bibnamefont
  {Mele}},\ }\bibfield  {title} {\bibinfo {title} {$z_2$ topological order and
  the quantum spin hall effect},\ }\href
  {https://doi.org/10.1103/physrevlett.95.146802} {\bibfield  {journal}
  {\bibinfo  {journal} {Phys. Rev. Lett.}\ }\textbf {\bibinfo {volume} {95}},\
  \bibinfo {pages} {146802} (\bibinfo {year} {2005})}\BibitemShut {NoStop}%
\bibitem [{\citenamefont {Bernevig}\ and\ \citenamefont
  {Zhang}(2006)}]{Bernevig2006}%
  \BibitemOpen
  \bibfield  {author} {\bibinfo {author} {\bibfnamefont {B.~A.}\ \bibnamefont
  {Bernevig}}\ and\ \bibinfo {author} {\bibfnamefont {S.-C.}\ \bibnamefont
  {Zhang}},\ }\bibfield  {title} {\bibinfo {title} {Quantum spin hall effect},\
  }\href {https://doi.org/10.1103/physrevlett.96.106802} {\bibfield  {journal}
  {\bibinfo  {journal} {Phys. Rev. Lett.}\ }\textbf {\bibinfo {volume} {96}},\
  \bibinfo {pages} {106802} (\bibinfo {year} {2006})}\BibitemShut {NoStop}%
\bibitem [{\citenamefont {König}\ \emph {et~al.}(2007)\citenamefont {König},
  \citenamefont {Wiedmann}, \citenamefont {Brune}, \citenamefont {Roth},
  \citenamefont {Buhmann}, \citenamefont {Molenkamp}, \citenamefont {Qi},\ and\
  \citenamefont {Zhang}}]{Koenig2007}%
  \BibitemOpen
  \bibfield  {author} {\bibinfo {author} {\bibfnamefont {M.}~\bibnamefont
  {König}}, \bibinfo {author} {\bibfnamefont {S.}~\bibnamefont {Wiedmann}},
  \bibinfo {author} {\bibfnamefont {C.}~\bibnamefont {Brune}}, \bibinfo
  {author} {\bibfnamefont {A.}~\bibnamefont {Roth}}, \bibinfo {author}
  {\bibfnamefont {H.}~\bibnamefont {Buhmann}}, \bibinfo {author} {\bibfnamefont
  {L.~W.}\ \bibnamefont {Molenkamp}}, \bibinfo {author} {\bibfnamefont {X.-L.}\
  \bibnamefont {Qi}},\ and\ \bibinfo {author} {\bibfnamefont {S.-C.}\
  \bibnamefont {Zhang}},\ }\bibfield  {title} {\bibinfo {title} {Quantum spin
  hall insulator state in {HgTe} quantum wells},\ }\href
  {https://doi.org/10.1126/science.1148047} {\bibfield  {journal} {\bibinfo
  {journal} {Science}\ }\textbf {\bibinfo {volume} {318}},\ \bibinfo {pages}
  {766} (\bibinfo {year} {2007})}\BibitemShut {NoStop}%
\bibitem [{\citenamefont {Karch}\ \emph {et~al.}(2011)\citenamefont {Karch},
  \citenamefont {Drexler}, \citenamefont {Olbrich}, \citenamefont
  {Fehrenbacher}, \citenamefont {Hirmer}, \citenamefont {Glazov}, \citenamefont
  {Tarasenko}, \citenamefont {Ivchenko}, \citenamefont {Birkner}, \citenamefont
  {Eroms}, \citenamefont {Weiss}, \citenamefont {Yakimova}, \citenamefont
  {Lara-Avila}, \citenamefont {Kubatkin}, \citenamefont {Ostler}, \citenamefont
  {Seyller},\ and\ \citenamefont {Ganichev}}]{Karch2011}%
  \BibitemOpen
  \bibfield  {author} {\bibinfo {author} {\bibfnamefont {J.}~\bibnamefont
  {Karch}}, \bibinfo {author} {\bibfnamefont {C.}~\bibnamefont {Drexler}},
  \bibinfo {author} {\bibfnamefont {P.}~\bibnamefont {Olbrich}}, \bibinfo
  {author} {\bibfnamefont {M.}~\bibnamefont {Fehrenbacher}}, \bibinfo {author}
  {\bibfnamefont {M.}~\bibnamefont {Hirmer}}, \bibinfo {author} {\bibfnamefont
  {M.~M.}\ \bibnamefont {Glazov}}, \bibinfo {author} {\bibfnamefont {S.~A.}\
  \bibnamefont {Tarasenko}}, \bibinfo {author} {\bibfnamefont {E.~L.}\
  \bibnamefont {Ivchenko}}, \bibinfo {author} {\bibfnamefont {B.}~\bibnamefont
  {Birkner}}, \bibinfo {author} {\bibfnamefont {J.}~\bibnamefont {Eroms}},
  \bibinfo {author} {\bibfnamefont {D.}~\bibnamefont {Weiss}}, \bibinfo
  {author} {\bibfnamefont {R.}~\bibnamefont {Yakimova}}, \bibinfo {author}
  {\bibfnamefont {S.}~\bibnamefont {Lara-Avila}}, \bibinfo {author}
  {\bibfnamefont {S.}~\bibnamefont {Kubatkin}}, \bibinfo {author}
  {\bibfnamefont {M.}~\bibnamefont {Ostler}}, \bibinfo {author} {\bibfnamefont
  {T.}~\bibnamefont {Seyller}},\ and\ \bibinfo {author} {\bibfnamefont {S.~D.}\
  \bibnamefont {Ganichev}},\ }\bibfield  {title} {\bibinfo {title} {Terahertz
  radiation driven chiral edge currents in graphene},\ }\href
  {https://doi.org/10.1103/physrevlett.107.276601} {\bibfield  {journal}
  {\bibinfo  {journal} {Phys. Rev. Lett.}\ }\textbf {\bibinfo {volume} {107}},\
  \bibinfo {pages} {276601} (\bibinfo {year} {2011})}\BibitemShut {NoStop}%
\bibitem [{\citenamefont {Otteneder}\ \emph {et~al.}(2020)\citenamefont
  {Otteneder}, \citenamefont {Hubmann}, \citenamefont {Lu}, \citenamefont
  {Kozlov}, \citenamefont {Golub}, \citenamefont {Watanabe}, \citenamefont
  {Taniguchi}, \citenamefont {Efetov},\ and\ \citenamefont
  {Ganichev}}]{Otteneder2020}%
  \BibitemOpen
  \bibfield  {author} {\bibinfo {author} {\bibfnamefont {M.}~\bibnamefont
  {Otteneder}}, \bibinfo {author} {\bibfnamefont {S.}~\bibnamefont {Hubmann}},
  \bibinfo {author} {\bibfnamefont {X.}~\bibnamefont {Lu}}, \bibinfo {author}
  {\bibfnamefont {D.~A.}\ \bibnamefont {Kozlov}}, \bibinfo {author}
  {\bibfnamefont {L.~E.}\ \bibnamefont {Golub}}, \bibinfo {author}
  {\bibfnamefont {K.}~\bibnamefont {Watanabe}}, \bibinfo {author}
  {\bibfnamefont {T.}~\bibnamefont {Taniguchi}}, \bibinfo {author}
  {\bibfnamefont {D.~K.}\ \bibnamefont {Efetov}},\ and\ \bibinfo {author}
  {\bibfnamefont {S.~D.}\ \bibnamefont {Ganichev}},\ }\bibfield  {title}
  {\bibinfo {title} {Terahertz photogalvanics in twisted bilayer graphene close
  to the second magic angle},\ }\href
  {https://doi.org/10.1021/acs.nanolett.0c02474} {\bibfield  {journal}
  {\bibinfo  {journal} {Nano Lett.}\ }\textbf {\bibinfo {volume} {20}},\
  \bibinfo {pages} {7152} (\bibinfo {year} {2020})}\BibitemShut {NoStop}%
\bibitem [{\citenamefont {Dantscher}\ \emph {et~al.}(2017)\citenamefont
  {Dantscher}, \citenamefont {Kozlov}, \citenamefont {Scherr}, \citenamefont
  {Gebert}, \citenamefont {Bärenfänger}, \citenamefont {Durnev},
  \citenamefont {Tarasenko}, \citenamefont {Bel'kov}, \citenamefont
  {Mikhailov}, \citenamefont {Dvoretsky}, \citenamefont {Kvon}, \citenamefont
  {Ziegler}, \citenamefont {Weiss},\ and\ \citenamefont
  {Ganichev}}]{Dantscher2017}%
  \BibitemOpen
  \bibfield  {author} {\bibinfo {author} {\bibfnamefont {K.-M.}\ \bibnamefont
  {Dantscher}}, \bibinfo {author} {\bibfnamefont {D.~A.}\ \bibnamefont
  {Kozlov}}, \bibinfo {author} {\bibfnamefont {M.~T.}\ \bibnamefont {Scherr}},
  \bibinfo {author} {\bibfnamefont {S.}~\bibnamefont {Gebert}}, \bibinfo
  {author} {\bibfnamefont {J.}~\bibnamefont {Bärenfänger}}, \bibinfo {author}
  {\bibfnamefont {M.~V.}\ \bibnamefont {Durnev}}, \bibinfo {author}
  {\bibfnamefont {S.~A.}\ \bibnamefont {Tarasenko}}, \bibinfo {author}
  {\bibfnamefont {V.~V.}\ \bibnamefont {Bel'kov}}, \bibinfo {author}
  {\bibfnamefont {N.~N.}\ \bibnamefont {Mikhailov}}, \bibinfo {author}
  {\bibfnamefont {S.~A.}\ \bibnamefont {Dvoretsky}}, \bibinfo {author}
  {\bibfnamefont {Z.~D.}\ \bibnamefont {Kvon}}, \bibinfo {author}
  {\bibfnamefont {J.}~\bibnamefont {Ziegler}}, \bibinfo {author} {\bibfnamefont
  {D.}~\bibnamefont {Weiss}},\ and\ \bibinfo {author} {\bibfnamefont {S.~D.}\
  \bibnamefont {Ganichev}},\ }\bibfield  {title} {\bibinfo {title}
  {Photogalvanic probing of helical edge channels in two-dimensional {HgTe}
  topological insulators},\ }\href {https://doi.org/10.1103/physrevb.95.201103}
  {\bibfield  {journal} {\bibinfo  {journal} {Phys. Rev. B}\ }\textbf {\bibinfo
  {volume} {95}},\ \bibinfo {pages} {201103(R)} (\bibinfo {year}
  {2017})}\BibitemShut {NoStop}%
\bibitem [{\citenamefont {Kiemle}\ \emph {et~al.}(2020)\citenamefont {Kiemle},
  \citenamefont {Seifert}, \citenamefont {Holleitner},\ and\ \citenamefont
  {Kastl}}]{Kiemle2021}%
  \BibitemOpen
  \bibfield  {author} {\bibinfo {author} {\bibfnamefont {J.}~\bibnamefont
  {Kiemle}}, \bibinfo {author} {\bibfnamefont {P.}~\bibnamefont {Seifert}},
  \bibinfo {author} {\bibfnamefont {A.~W.}\ \bibnamefont {Holleitner}},\ and\
  \bibinfo {author} {\bibfnamefont {C.}~\bibnamefont {Kastl}},\ }\bibfield
  {title} {\bibinfo {title} {Ultrafast and local optoelectronic transport in
  topological insulators},\ }\href {https://doi.org/10.1002/pssb.202000033}
  {\bibfield  {journal} {\bibinfo  {journal} {physica status solidi (b)}\
  }\textbf {\bibinfo {volume} {258}},\ \bibinfo {pages} {2000033} (\bibinfo
  {year} {2020})}\BibitemShut {NoStop}%
\bibitem [{\citenamefont {Glazov}\ and\ \citenamefont
  {Ganichev}(2014)}]{GlazovGanichev_review}%
  \BibitemOpen
  \bibfield  {author} {\bibinfo {author} {\bibfnamefont {M.}~\bibnamefont
  {Glazov}}\ and\ \bibinfo {author} {\bibfnamefont {S.}~\bibnamefont
  {Ganichev}},\ }\bibfield  {title} {\bibinfo {title} {High frequency electric
  field induced nonlinear effects in graphene},\ }\href
  {https://doi.org/10.1016/j.physrep.2013.10.003} {\bibfield  {journal}
  {\bibinfo  {journal} {Phys. Rep.}\ }\textbf {\bibinfo {volume} {535}},\
  \bibinfo {pages} {101} (\bibinfo {year} {2014})}\BibitemShut {NoStop}%
\bibitem [{\citenamefont {Magarill}\ and\ \citenamefont
  {Entin}(2015)}]{Magarill2015}%
  \BibitemOpen
  \bibfield  {author} {\bibinfo {author} {\bibfnamefont {L.~I.}\ \bibnamefont
  {Magarill}}\ and\ \bibinfo {author} {\bibfnamefont {M.~V.}\ \bibnamefont
  {Entin}},\ }\bibfield  {title} {\bibinfo {title} {Surface photocurrent in an
  electron gas over liquid he subjected to a quantizing magnetic field},\
  }\href {https://doi.org/10.1134/s0021364015110090} {\bibfield  {journal}
  {\bibinfo  {journal} {{JETP} Letters}\ }\textbf {\bibinfo {volume} {101}},\
  \bibinfo {pages} {744} (\bibinfo {year} {2015})}\BibitemShut {NoStop}%
\bibitem [{\citenamefont {Plank}\ \emph {et~al.}(2018)\citenamefont {Plank},
  \citenamefont {Durnev}, \citenamefont {Candussio}, \citenamefont {Pernul},
  \citenamefont {Dantscher}, \citenamefont {Mönch}, \citenamefont {Sandner},
  \citenamefont {Eroms}, \citenamefont {Weiss}, \citenamefont {Bel'kov},
  \citenamefont {Tarasenko},\ and\ \citenamefont {Ganichev}}]{Plank2019}%
  \BibitemOpen
  \bibfield  {author} {\bibinfo {author} {\bibfnamefont {H.}~\bibnamefont
  {Plank}}, \bibinfo {author} {\bibfnamefont {M.~V.}\ \bibnamefont {Durnev}},
  \bibinfo {author} {\bibfnamefont {S.}~\bibnamefont {Candussio}}, \bibinfo
  {author} {\bibfnamefont {J.}~\bibnamefont {Pernul}}, \bibinfo {author}
  {\bibfnamefont {K.-M.}\ \bibnamefont {Dantscher}}, \bibinfo {author}
  {\bibfnamefont {E.}~\bibnamefont {Mönch}}, \bibinfo {author} {\bibfnamefont
  {A.}~\bibnamefont {Sandner}}, \bibinfo {author} {\bibfnamefont
  {J.}~\bibnamefont {Eroms}}, \bibinfo {author} {\bibfnamefont
  {D.}~\bibnamefont {Weiss}}, \bibinfo {author} {\bibfnamefont {V.~V.}\
  \bibnamefont {Bel'kov}}, \bibinfo {author} {\bibfnamefont {S.~A.}\
  \bibnamefont {Tarasenko}},\ and\ \bibinfo {author} {\bibfnamefont {S.~D.}\
  \bibnamefont {Ganichev}},\ }\bibfield  {title} {\bibinfo {title} {Edge
  currents driven by terahertz radiation in graphene in quantum hall regime},\
  }\href {https://doi.org/10.1088/2053-1583/aae39c} {\bibfield  {journal}
  {\bibinfo  {journal} {2D Mater.}\ }\textbf {\bibinfo {volume} {6}},\ \bibinfo
  {pages} {011002} (\bibinfo {year} {2018})}\BibitemShut {NoStop}%
\bibitem [{\citenamefont {Durnev}\ and\ \citenamefont
  {Tarasenko}(2019)}]{Durnev2019}%
  \BibitemOpen
  \bibfield  {author} {\bibinfo {author} {\bibfnamefont {M.~V.}\ \bibnamefont
  {Durnev}}\ and\ \bibinfo {author} {\bibfnamefont {S.~A.}\ \bibnamefont
  {Tarasenko}},\ }\bibfield  {title} {\bibinfo {title} {High-frequency
  nonlinear transport and photogalvanic effects in 2d topological insulators},\
  }\href {https://doi.org/10.1002/andp.201800418} {\bibfield  {journal}
  {\bibinfo  {journal} {Ann. Phys.}\ }\textbf {\bibinfo {volume} {531}},\
  \bibinfo {pages} {1800418} (\bibinfo {year} {2019})}\BibitemShut {NoStop}%
\bibitem [{\citenamefont {Ma}\ \emph {et~al.}(2018)\citenamefont {Ma},
  \citenamefont {Lui}, \citenamefont {Song}, \citenamefont {Lin}, \citenamefont
  {Kong}, \citenamefont {Cao}, \citenamefont {Dinh}, \citenamefont {Nair},
  \citenamefont {Fang}, \citenamefont {Watanabe}, \citenamefont {Taniguchi},
  \citenamefont {Xu}, \citenamefont {Kong}, \citenamefont {Palacios},
  \citenamefont {Gedik}, \citenamefont {Gabor},\ and\ \citenamefont
  {Jarillo-Herrero}}]{Ma2019}%
  \BibitemOpen
  \bibfield  {author} {\bibinfo {author} {\bibfnamefont {Q.}~\bibnamefont
  {Ma}}, \bibinfo {author} {\bibfnamefont {C.~H.}\ \bibnamefont {Lui}},
  \bibinfo {author} {\bibfnamefont {J.~C.~W.}\ \bibnamefont {Song}}, \bibinfo
  {author} {\bibfnamefont {Y.}~\bibnamefont {Lin}}, \bibinfo {author}
  {\bibfnamefont {J.~F.}\ \bibnamefont {Kong}}, \bibinfo {author}
  {\bibfnamefont {Y.}~\bibnamefont {Cao}}, \bibinfo {author} {\bibfnamefont
  {T.~H.}\ \bibnamefont {Dinh}}, \bibinfo {author} {\bibfnamefont {N.~L.}\
  \bibnamefont {Nair}}, \bibinfo {author} {\bibfnamefont {W.}~\bibnamefont
  {Fang}}, \bibinfo {author} {\bibfnamefont {K.}~\bibnamefont {Watanabe}},
  \bibinfo {author} {\bibfnamefont {T.}~\bibnamefont {Taniguchi}}, \bibinfo
  {author} {\bibfnamefont {S.-Y.}\ \bibnamefont {Xu}}, \bibinfo {author}
  {\bibfnamefont {J.}~\bibnamefont {Kong}}, \bibinfo {author} {\bibfnamefont
  {T.}~\bibnamefont {Palacios}}, \bibinfo {author} {\bibfnamefont
  {N.}~\bibnamefont {Gedik}}, \bibinfo {author} {\bibfnamefont {N.~M.}\
  \bibnamefont {Gabor}},\ and\ \bibinfo {author} {\bibfnamefont
  {P.}~\bibnamefont {Jarillo-Herrero}},\ }\bibfield  {title} {\bibinfo {title}
  {Giant intrinsic photoresponse in pristine graphene},\ }\href
  {https://doi.org/10.1038/s41565-018-0323-8} {\bibfield  {journal} {\bibinfo
  {journal} {Nat. Nanotechnol.}\ }\textbf {\bibinfo {volume} {14}},\ \bibinfo
  {pages} {145} (\bibinfo {year} {2018})}\BibitemShut {NoStop}%
\bibitem [{\citenamefont {Wang}\ \emph {et~al.}(2019)\citenamefont {Wang},
  \citenamefont {Zheng}, \citenamefont {He}, \citenamefont {Cao}, \citenamefont
  {Liu}, \citenamefont {Wang}, \citenamefont {Ma}, \citenamefont {Lai},
  \citenamefont {Lu}, \citenamefont {Jia}, \citenamefont {Yan}, \citenamefont
  {Shi}, \citenamefont {Duan}, \citenamefont {Han}, \citenamefont {Xiao},
  \citenamefont {Chen}, \citenamefont {Sun}, \citenamefont {Yao},\ and\
  \citenamefont {Sun}}]{Wang2019}%
  \BibitemOpen
  \bibfield  {author} {\bibinfo {author} {\bibfnamefont {Q.}~\bibnamefont
  {Wang}}, \bibinfo {author} {\bibfnamefont {J.}~\bibnamefont {Zheng}},
  \bibinfo {author} {\bibfnamefont {Y.}~\bibnamefont {He}}, \bibinfo {author}
  {\bibfnamefont {J.}~\bibnamefont {Cao}}, \bibinfo {author} {\bibfnamefont
  {X.}~\bibnamefont {Liu}}, \bibinfo {author} {\bibfnamefont {M.}~\bibnamefont
  {Wang}}, \bibinfo {author} {\bibfnamefont {J.}~\bibnamefont {Ma}}, \bibinfo
  {author} {\bibfnamefont {J.}~\bibnamefont {Lai}}, \bibinfo {author}
  {\bibfnamefont {H.}~\bibnamefont {Lu}}, \bibinfo {author} {\bibfnamefont
  {S.}~\bibnamefont {Jia}}, \bibinfo {author} {\bibfnamefont {D.}~\bibnamefont
  {Yan}}, \bibinfo {author} {\bibfnamefont {Y.}~\bibnamefont {Shi}}, \bibinfo
  {author} {\bibfnamefont {J.}~\bibnamefont {Duan}}, \bibinfo {author}
  {\bibfnamefont {J.}~\bibnamefont {Han}}, \bibinfo {author} {\bibfnamefont
  {W.}~\bibnamefont {Xiao}}, \bibinfo {author} {\bibfnamefont {J.-H.}\
  \bibnamefont {Chen}}, \bibinfo {author} {\bibfnamefont {K.}~\bibnamefont
  {Sun}}, \bibinfo {author} {\bibfnamefont {Y.}~\bibnamefont {Yao}},\ and\
  \bibinfo {author} {\bibfnamefont {D.}~\bibnamefont {Sun}},\ }\bibfield
  {title} {\bibinfo {title} {Robust edge photocurrent response on layered type
  {II} weyl semimetal {WTe}2},\ }\href
  {https://doi.org/10.1038/s41467-019-13713-1} {\bibfield  {journal} {\bibinfo
  {journal} {Nat. Commun.}\ }\textbf {\bibinfo {volume} {10}},\ \bibinfo
  {pages} {5736} (\bibinfo {year} {2019})}\BibitemShut {NoStop}%
\bibitem [{\citenamefont {Candussio}\ \emph {et~al.}(2020)\citenamefont
  {Candussio}, \citenamefont {Durnev}, \citenamefont {Tarasenko}, \citenamefont
  {Yin}, \citenamefont {Keil}, \citenamefont {Yang}, \citenamefont {Son},
  \citenamefont {Mishchenko}, \citenamefont {Plank}, \citenamefont {Bel'kov},
  \citenamefont {Slizovskiy}, \citenamefont {Fal'ko},\ and\ \citenamefont
  {Ganichev}}]{Candussio2020}%
  \BibitemOpen
  \bibfield  {author} {\bibinfo {author} {\bibfnamefont {S.}~\bibnamefont
  {Candussio}}, \bibinfo {author} {\bibfnamefont {M.~V.}\ \bibnamefont
  {Durnev}}, \bibinfo {author} {\bibfnamefont {S.~A.}\ \bibnamefont
  {Tarasenko}}, \bibinfo {author} {\bibfnamefont {J.}~\bibnamefont {Yin}},
  \bibinfo {author} {\bibfnamefont {J.}~\bibnamefont {Keil}}, \bibinfo {author}
  {\bibfnamefont {Y.}~\bibnamefont {Yang}}, \bibinfo {author} {\bibfnamefont
  {S.-K.}\ \bibnamefont {Son}}, \bibinfo {author} {\bibfnamefont
  {A.}~\bibnamefont {Mishchenko}}, \bibinfo {author} {\bibfnamefont
  {H.}~\bibnamefont {Plank}}, \bibinfo {author} {\bibfnamefont {V.~V.}\
  \bibnamefont {Bel'kov}}, \bibinfo {author} {\bibfnamefont {S.}~\bibnamefont
  {Slizovskiy}}, \bibinfo {author} {\bibfnamefont {V.}~\bibnamefont {Fal'ko}},\
  and\ \bibinfo {author} {\bibfnamefont {S.~D.}\ \bibnamefont {Ganichev}},\
  }\bibfield  {title} {\bibinfo {title} {Edge photocurrent driven by terahertz
  electric field in bilayer graphene},\ }\href
  {https://doi.org/10.1103/physrevb.102.045406} {\bibfield  {journal} {\bibinfo
   {journal} {Phys. Rev. B}\ }\textbf {\bibinfo {volume} {102}},\ \bibinfo
  {pages} {045406} (\bibinfo {year} {2020})}\BibitemShut {NoStop}%
\bibitem [{\citenamefont {Durnev}\ and\ \citenamefont
  {Tarasenko}(2020)}]{Durnev_pssb2021}%
  \BibitemOpen
  \bibfield  {author} {\bibinfo {author} {\bibfnamefont {M.~V.}\ \bibnamefont
  {Durnev}}\ and\ \bibinfo {author} {\bibfnamefont {S.~A.}\ \bibnamefont
  {Tarasenko}},\ }\bibfield  {title} {\bibinfo {title} {Rectification of {AC}
  electric current at the edge of 2d electron gas},\ }\href
  {https://doi.org/10.1002/pssb.202000291} {\bibfield  {journal} {\bibinfo
  {journal} {physica status solidi (b)}\ }\textbf {\bibinfo {volume} {258}},\
  \bibinfo {pages} {2000291} (\bibinfo {year} {2020})}\BibitemShut {NoStop}%
\bibitem [{\citenamefont {Durnev}\ and\ \citenamefont
  {Tarasenko}(2021)}]{Durnev2021PRB}%
  \BibitemOpen
  \bibfield  {author} {\bibinfo {author} {\bibfnamefont {M.~V.}\ \bibnamefont
  {Durnev}}\ and\ \bibinfo {author} {\bibfnamefont {S.~A.}\ \bibnamefont
  {Tarasenko}},\ }\bibfield  {title} {\bibinfo {title} {Edge photogalvanic
  effect caused by optical alignment of carrier momenta in two-dimensional
  dirac materials},\ }\href {https://doi.org/10.1103/physrevb.103.165411}
  {\bibfield  {journal} {\bibinfo  {journal} {Phys. Rev. B}\ }\textbf {\bibinfo
  {volume} {103}},\ \bibinfo {pages} {165411} (\bibinfo {year}
  {2021})}\BibitemShut {NoStop}%
\bibitem [{\citenamefont {Candussio}\ \emph {et~al.}(2021)\citenamefont
  {Candussio}, \citenamefont {Durnev}, \citenamefont {Slizovskiy},
  \citenamefont {Jötten}, \citenamefont {Keil}, \citenamefont {Bel'kov},
  \citenamefont {Yin}, \citenamefont {Yang}, \citenamefont {Son}, \citenamefont
  {Mishchenko}, \citenamefont {Fal'ko},\ and\ \citenamefont
  {Ganichev}}]{Candussio2021}%
  \BibitemOpen
  \bibfield  {author} {\bibinfo {author} {\bibfnamefont {S.}~\bibnamefont
  {Candussio}}, \bibinfo {author} {\bibfnamefont {M.~V.}\ \bibnamefont
  {Durnev}}, \bibinfo {author} {\bibfnamefont {S.}~\bibnamefont {Slizovskiy}},
  \bibinfo {author} {\bibfnamefont {T.}~\bibnamefont {Jötten}}, \bibinfo
  {author} {\bibfnamefont {J.}~\bibnamefont {Keil}}, \bibinfo {author}
  {\bibfnamefont {V.~V.}\ \bibnamefont {Bel'kov}}, \bibinfo {author}
  {\bibfnamefont {J.}~\bibnamefont {Yin}}, \bibinfo {author} {\bibfnamefont
  {Y.}~\bibnamefont {Yang}}, \bibinfo {author} {\bibfnamefont {S.-K.}\
  \bibnamefont {Son}}, \bibinfo {author} {\bibfnamefont {A.}~\bibnamefont
  {Mishchenko}}, \bibinfo {author} {\bibfnamefont {V.}~\bibnamefont {Fal'ko}},\
  and\ \bibinfo {author} {\bibfnamefont {S.~D.}\ \bibnamefont {Ganichev}},\
  }\bibfield  {title} {\bibinfo {title} {Edge photocurrent in bilayer graphene
  due to inter-landau-level transitions},\ }\href
  {https://doi.org/10.1103/physrevb.103.125408} {\bibfield  {journal} {\bibinfo
   {journal} {Phys. Rev. B}\ }\textbf {\bibinfo {volume} {103}},\ \bibinfo
  {pages} {125408} (\bibinfo {year} {2021})}\BibitemShut {NoStop}%
\bibitem [{\citenamefont {Ivchenko}\ and\ \citenamefont
  {Ganichev}(2018)}]{Ivchenko2018}%
  \BibitemOpen
  \bibfield  {author} {\bibinfo {author} {\bibfnamefont {E.}~\bibnamefont
  {Ivchenko}}\ and\ \bibinfo {author} {\bibfnamefont {S.}~\bibnamefont
  {Ganichev}},\ }\href@noop {} {\emph {\bibinfo {title} {Spin Physics in
  Semiconductors}}},\ edited by\ \bibinfo {editor} {\bibfnamefont
  {M.}~\bibnamefont {Dyakonov}}\ (\bibinfo  {publisher} {Springer},\ \bibinfo
  {year} {2018})\ \bibinfo {note} {281-328}\BibitemShut {NoStop}%
\bibitem [{\citenamefont {Mics}\ \emph {et~al.}(2015)\citenamefont {Mics},
  \citenamefont {Tielrooij}, \citenamefont {Parvez}, \citenamefont {Jensen},
  \citenamefont {Ivanov}, \citenamefont {Feng}, \citenamefont {Müllen},
  \citenamefont {Bonn},\ and\ \citenamefont {Turchinovich}}]{Mics2015}%
  \BibitemOpen
  \bibfield  {author} {\bibinfo {author} {\bibfnamefont {Z.}~\bibnamefont
  {Mics}}, \bibinfo {author} {\bibfnamefont {K.-J.}\ \bibnamefont {Tielrooij}},
  \bibinfo {author} {\bibfnamefont {K.}~\bibnamefont {Parvez}}, \bibinfo
  {author} {\bibfnamefont {S.~A.}\ \bibnamefont {Jensen}}, \bibinfo {author}
  {\bibfnamefont {I.}~\bibnamefont {Ivanov}}, \bibinfo {author} {\bibfnamefont
  {X.}~\bibnamefont {Feng}}, \bibinfo {author} {\bibfnamefont {K.}~\bibnamefont
  {Müllen}}, \bibinfo {author} {\bibfnamefont {M.}~\bibnamefont {Bonn}},\ and\
  \bibinfo {author} {\bibfnamefont {D.}~\bibnamefont {Turchinovich}},\
  }\bibfield  {title} {\bibinfo {title} {Thermodynamic picture of ultrafast
  charge transport in graphene},\ }\href {https://doi.org/10.1038/ncomms8655}
  {\bibfield  {journal} {\bibinfo  {journal} {Nat. Commun.}\ }\textbf {\bibinfo
  {volume} {6}},\ \bibinfo {pages} {7655} (\bibinfo {year} {2015})}\BibitemShut
  {NoStop}%
\bibitem [{\citenamefont {Bao}\ \emph {et~al.}(2009)\citenamefont {Bao},
  \citenamefont {Zhang}, \citenamefont {Wang}, \citenamefont {Ni},
  \citenamefont {Yan}, \citenamefont {Shen}, \citenamefont {Loh},\ and\
  \citenamefont {Tang}}]{Bao2009}%
  \BibitemOpen
  \bibfield  {author} {\bibinfo {author} {\bibfnamefont {Q.}~\bibnamefont
  {Bao}}, \bibinfo {author} {\bibfnamefont {H.}~\bibnamefont {Zhang}}, \bibinfo
  {author} {\bibfnamefont {Y.}~\bibnamefont {Wang}}, \bibinfo {author}
  {\bibfnamefont {Z.}~\bibnamefont {Ni}}, \bibinfo {author} {\bibfnamefont
  {Y.}~\bibnamefont {Yan}}, \bibinfo {author} {\bibfnamefont {Z.~X.}\
  \bibnamefont {Shen}}, \bibinfo {author} {\bibfnamefont {K.~P.}\ \bibnamefont
  {Loh}},\ and\ \bibinfo {author} {\bibfnamefont {D.~Y.}\ \bibnamefont
  {Tang}},\ }\bibfield  {title} {\bibinfo {title} {Atomic-layer graphene as a
  saturable absorber for ultrafast pulsed lasers},\ }\href
  {https://doi.org/10.1002/adfm.200901007} {\bibfield  {journal} {\bibinfo
  {journal} {Adv. Funct. Mater.}\ }\textbf {\bibinfo {volume} {19}},\ \bibinfo
  {pages} {3077} (\bibinfo {year} {2009})}\BibitemShut {NoStop}%
\bibitem [{\citenamefont {Kumar}\ \emph {et~al.}(2009)\citenamefont {Kumar},
  \citenamefont {Anija}, \citenamefont {Kamaraju}, \citenamefont {Vasu},
  \citenamefont {Subrahmanyam}, \citenamefont {Sood},\ and\ \citenamefont
  {Rao}}]{Kumar2009}%
  \BibitemOpen
  \bibfield  {author} {\bibinfo {author} {\bibfnamefont {S.}~\bibnamefont
  {Kumar}}, \bibinfo {author} {\bibfnamefont {M.}~\bibnamefont {Anija}},
  \bibinfo {author} {\bibfnamefont {N.}~\bibnamefont {Kamaraju}}, \bibinfo
  {author} {\bibfnamefont {K.~S.}\ \bibnamefont {Vasu}}, \bibinfo {author}
  {\bibfnamefont {K.~S.}\ \bibnamefont {Subrahmanyam}}, \bibinfo {author}
  {\bibfnamefont {A.~K.}\ \bibnamefont {Sood}},\ and\ \bibinfo {author}
  {\bibfnamefont {C.~N.~R.}\ \bibnamefont {Rao}},\ }\bibfield  {title}
  {\bibinfo {title} {Femtosecond carrier dynamics and saturable absorption in
  graphene suspensions},\ }\href {https://doi.org/10.1063/1.3264964} {\bibfield
   {journal} {\bibinfo  {journal} {Appl. Phys. Lett.}\ }\textbf {\bibinfo
  {volume} {95}},\ \bibinfo {pages} {191911} (\bibinfo {year}
  {2009})}\BibitemShut {NoStop}%
\bibitem [{\citenamefont {Vasko}(2010)}]{Vasko2010}%
  \BibitemOpen
  \bibfield  {author} {\bibinfo {author} {\bibfnamefont {F.~T.}\ \bibnamefont
  {Vasko}},\ }\bibfield  {title} {\bibinfo {title} {Saturation of interband
  absorption in graphene},\ }\href {https://doi.org/10.1103/physrevb.82.245422}
  {\bibfield  {journal} {\bibinfo  {journal} {Phys. Rev. B}\ }\textbf {\bibinfo
  {volume} {82}},\ \bibinfo {pages} {245422} (\bibinfo {year}
  {2010})}\BibitemShut {NoStop}%
\bibitem [{\citenamefont {Weis}\ \emph {et~al.}(2012)\citenamefont {Weis},
  \citenamefont {Garcia-Pomar}, \citenamefont {Höh}, \citenamefont {Reinhard},
  \citenamefont {Brodyanski},\ and\ \citenamefont {Rahm}}]{Weis2012}%
  \BibitemOpen
  \bibfield  {author} {\bibinfo {author} {\bibfnamefont {P.}~\bibnamefont
  {Weis}}, \bibinfo {author} {\bibfnamefont {J.~L.}\ \bibnamefont
  {Garcia-Pomar}}, \bibinfo {author} {\bibfnamefont {M.}~\bibnamefont {Höh}},
  \bibinfo {author} {\bibfnamefont {B.}~\bibnamefont {Reinhard}}, \bibinfo
  {author} {\bibfnamefont {A.}~\bibnamefont {Brodyanski}},\ and\ \bibinfo
  {author} {\bibfnamefont {M.}~\bibnamefont {Rahm}},\ }\bibfield  {title}
  {\bibinfo {title} {Spectrally wide-band terahertz wave modulator based on
  optically tuned graphene},\ }\href {https://doi.org/10.1021/nn303392s}
  {\bibfield  {journal} {\bibinfo  {journal} {{ACS} Nano}\ }\textbf {\bibinfo
  {volume} {6}},\ \bibinfo {pages} {9118} (\bibinfo {year} {2012})}\BibitemShut
  {NoStop}%
\bibitem [{\citenamefont {Avouris}\ and\ \citenamefont
  {Xia}(2012)}]{Avouris2012}%
  \BibitemOpen
  \bibfield  {author} {\bibinfo {author} {\bibfnamefont {P.}~\bibnamefont
  {Avouris}}\ and\ \bibinfo {author} {\bibfnamefont {F.}~\bibnamefont {Xia}},\
  }\bibfield  {title} {\bibinfo {title} {Graphene applications in electronics
  and photonics},\ }\href {https://doi.org/10.1557/mrs.2012.206} {\bibfield
  {journal} {\bibinfo  {journal} {{MRS} Bulletin}\ }\textbf {\bibinfo {volume}
  {37}},\ \bibinfo {pages} {1225} (\bibinfo {year} {2012})}\BibitemShut
  {NoStop}%
\bibitem [{\citenamefont {Winnerl}\ \emph {et~al.}(2013)\citenamefont
  {Winnerl}, \citenamefont {Göttfert}, \citenamefont {Mittendorff},
  \citenamefont {Schneider}, \citenamefont {Helm}, \citenamefont {Winzer},
  \citenamefont {Malic}, \citenamefont {Knorr}, \citenamefont {Orlita},
  \citenamefont {Potemski}, \citenamefont {Sprinkle}, \citenamefont {Berger},\
  and\ \citenamefont {de~Heer}}]{Winnerl2013}%
  \BibitemOpen
  \bibfield  {author} {\bibinfo {author} {\bibfnamefont {S.}~\bibnamefont
  {Winnerl}}, \bibinfo {author} {\bibfnamefont {F.}~\bibnamefont {Göttfert}},
  \bibinfo {author} {\bibfnamefont {M.}~\bibnamefont {Mittendorff}}, \bibinfo
  {author} {\bibfnamefont {H.}~\bibnamefont {Schneider}}, \bibinfo {author}
  {\bibfnamefont {M.}~\bibnamefont {Helm}}, \bibinfo {author} {\bibfnamefont
  {T.}~\bibnamefont {Winzer}}, \bibinfo {author} {\bibfnamefont
  {E.}~\bibnamefont {Malic}}, \bibinfo {author} {\bibfnamefont
  {A.}~\bibnamefont {Knorr}}, \bibinfo {author} {\bibfnamefont
  {M.}~\bibnamefont {Orlita}}, \bibinfo {author} {\bibfnamefont
  {M.}~\bibnamefont {Potemski}}, \bibinfo {author} {\bibfnamefont
  {M.}~\bibnamefont {Sprinkle}}, \bibinfo {author} {\bibfnamefont
  {C.}~\bibnamefont {Berger}},\ and\ \bibinfo {author} {\bibfnamefont {W.~A.}\
  \bibnamefont {de~Heer}},\ }\bibfield  {title} {\bibinfo {title}
  {Time-resolved spectroscopy on epitaxial graphene in the infrared spectral
  range: relaxation dynamics and saturation behavior},\ }\href
  {https://doi.org/10.1088/0953-8984/25/5/054202} {\bibfield  {journal}
  {\bibinfo  {journal} {J. Phys.: Condens. Matter}\ }\textbf {\bibinfo {volume}
  {25}},\ \bibinfo {pages} {054202} (\bibinfo {year} {2013})}\BibitemShut
  {NoStop}%
\bibitem [{\citenamefont {Mikhailov}(2017)}]{Michailov2017}%
  \BibitemOpen
  \bibfield  {author} {\bibinfo {author} {\bibfnamefont {S.~A.}\ \bibnamefont
  {Mikhailov}},\ }\bibfield  {title} {\bibinfo {title} {Nonperturbative
  quasiclassical theory of the nonlinear electrodynamic response of graphene},\
  }\href {https://doi.org/10.1103/physrevb.95.085432} {\bibfield  {journal}
  {\bibinfo  {journal} {Phys. Rev. B}\ }\textbf {\bibinfo {volume} {95}},\
  \bibinfo {pages} {085432} (\bibinfo {year} {2017})}\BibitemShut {NoStop}%
\bibitem [{\citenamefont {Bianchi}\ \emph {et~al.}(2017)\citenamefont
  {Bianchi}, \citenamefont {Carey}, \citenamefont {Viti}, \citenamefont {Li},
  \citenamefont {Linfield}, \citenamefont {Davies}, \citenamefont {Tredicucci},
  \citenamefont {Yoon}, \citenamefont {Karagiannidis}, \citenamefont
  {Lombardi}, \citenamefont {Tomarchio}, \citenamefont {Ferrari}, \citenamefont
  {Torrisi},\ and\ \citenamefont {Vitiello}}]{Bianchi2017}%
  \BibitemOpen
  \bibfield  {author} {\bibinfo {author} {\bibfnamefont {V.}~\bibnamefont
  {Bianchi}}, \bibinfo {author} {\bibfnamefont {T.}~\bibnamefont {Carey}},
  \bibinfo {author} {\bibfnamefont {L.}~\bibnamefont {Viti}}, \bibinfo {author}
  {\bibfnamefont {L.}~\bibnamefont {Li}}, \bibinfo {author} {\bibfnamefont
  {E.~H.}\ \bibnamefont {Linfield}}, \bibinfo {author} {\bibfnamefont {A.~G.}\
  \bibnamefont {Davies}}, \bibinfo {author} {\bibfnamefont {A.}~\bibnamefont
  {Tredicucci}}, \bibinfo {author} {\bibfnamefont {D.}~\bibnamefont {Yoon}},
  \bibinfo {author} {\bibfnamefont {P.~G.}\ \bibnamefont {Karagiannidis}},
  \bibinfo {author} {\bibfnamefont {L.}~\bibnamefont {Lombardi}}, \bibinfo
  {author} {\bibfnamefont {F.}~\bibnamefont {Tomarchio}}, \bibinfo {author}
  {\bibfnamefont {A.~C.}\ \bibnamefont {Ferrari}}, \bibinfo {author}
  {\bibfnamefont {F.}~\bibnamefont {Torrisi}},\ and\ \bibinfo {author}
  {\bibfnamefont {M.~S.}\ \bibnamefont {Vitiello}},\ }\bibfield  {title}
  {\bibinfo {title} {Terahertz saturable absorbers from liquid phase
  exfoliation of graphite},\ }\href {https://doi.org/10.1038/ncomms15763}
  {\bibfield  {journal} {\bibinfo  {journal} {Nat. Commun.}\ }\textbf {\bibinfo
  {volume} {8}},\ \bibinfo {pages} {15763} (\bibinfo {year}
  {2017})}\BibitemShut {NoStop}%
\bibitem [{\citenamefont {Autere}\ \emph {et~al.}(2018)\citenamefont {Autere},
  \citenamefont {Jussila}, \citenamefont {Dai}, \citenamefont {Wang},
  \citenamefont {Lipsanen},\ and\ \citenamefont {Sun}}]{Autere2018}%
  \BibitemOpen
  \bibfield  {author} {\bibinfo {author} {\bibfnamefont {A.}~\bibnamefont
  {Autere}}, \bibinfo {author} {\bibfnamefont {H.}~\bibnamefont {Jussila}},
  \bibinfo {author} {\bibfnamefont {Y.}~\bibnamefont {Dai}}, \bibinfo {author}
  {\bibfnamefont {Y.}~\bibnamefont {Wang}}, \bibinfo {author} {\bibfnamefont
  {H.}~\bibnamefont {Lipsanen}},\ and\ \bibinfo {author} {\bibfnamefont
  {Z.}~\bibnamefont {Sun}},\ }\bibfield  {title} {\bibinfo {title} {Nonlinear
  optics with 2d layered materials},\ }\href
  {https://doi.org/10.1002/adma.201705963} {\bibfield  {journal} {\bibinfo
  {journal} {Adv. Mater.}\ }\textbf {\bibinfo {volume} {30}},\ \bibinfo {pages}
  {1705963} (\bibinfo {year} {2018})}\BibitemShut {NoStop}%
\bibitem [{\citenamefont {Baudisch}\ \emph {et~al.}(2018)\citenamefont
  {Baudisch}, \citenamefont {Marini}, \citenamefont {Cox}, \citenamefont {Zhu},
  \citenamefont {Silva}, \citenamefont {Teichmann}, \citenamefont {Massicotte},
  \citenamefont {Koppens}, \citenamefont {Levitov}, \citenamefont {de~Abajo},\
  and\ \citenamefont {Biegert}}]{Baudisch2018}%
  \BibitemOpen
  \bibfield  {author} {\bibinfo {author} {\bibfnamefont {M.}~\bibnamefont
  {Baudisch}}, \bibinfo {author} {\bibfnamefont {A.}~\bibnamefont {Marini}},
  \bibinfo {author} {\bibfnamefont {J.~D.}\ \bibnamefont {Cox}}, \bibinfo
  {author} {\bibfnamefont {T.}~\bibnamefont {Zhu}}, \bibinfo {author}
  {\bibfnamefont {F.}~\bibnamefont {Silva}}, \bibinfo {author} {\bibfnamefont
  {S.}~\bibnamefont {Teichmann}}, \bibinfo {author} {\bibfnamefont
  {M.}~\bibnamefont {Massicotte}}, \bibinfo {author} {\bibfnamefont
  {F.}~\bibnamefont {Koppens}}, \bibinfo {author} {\bibfnamefont {L.~S.}\
  \bibnamefont {Levitov}}, \bibinfo {author} {\bibfnamefont {F.~J.~G.}\
  \bibnamefont {de~Abajo}},\ and\ \bibinfo {author} {\bibfnamefont
  {J.}~\bibnamefont {Biegert}},\ }\bibfield  {title} {\bibinfo {title}
  {Ultrafast nonlinear optical response of dirac fermions in graphene},\ }\href
  {https://doi.org/10.1038/s41467-018-03413-7} {\bibfield  {journal} {\bibinfo
  {journal} {Nat. Commun.}\ }\textbf {\bibinfo {volume} {9}},\ \bibinfo {pages}
  {1018} (\bibinfo {year} {2018})}\BibitemShut {NoStop}%
\bibitem [{\citenamefont {Yumoto}\ \emph {et~al.}(2018)\citenamefont {Yumoto},
  \citenamefont {Matsunaga}, \citenamefont {Hibino},\ and\ \citenamefont
  {Shimano}}]{Yumoto2018}%
  \BibitemOpen
  \bibfield  {author} {\bibinfo {author} {\bibfnamefont {G.}~\bibnamefont
  {Yumoto}}, \bibinfo {author} {\bibfnamefont {R.}~\bibnamefont {Matsunaga}},
  \bibinfo {author} {\bibfnamefont {H.}~\bibnamefont {Hibino}},\ and\ \bibinfo
  {author} {\bibfnamefont {R.}~\bibnamefont {Shimano}},\ }\bibfield  {title}
  {\bibinfo {title} {Ultrafast terahertz nonlinear optics of landau level
  transitions in a monolayer graphene},\ }\href
  {https://doi.org/10.1103/physrevlett.120.107401} {\bibfield  {journal}
  {\bibinfo  {journal} {Phys. Rev. Lett.}\ }\textbf {\bibinfo {volume} {120}},\
  \bibinfo {pages} {107401} (\bibinfo {year} {2018})}\BibitemShut {NoStop}%
\bibitem [{\citenamefont {You}\ \emph {et~al.}(2018)\citenamefont {You},
  \citenamefont {Bongu}, \citenamefont {Bao},\ and\ \citenamefont
  {Panoiu}}]{You2019}%
  \BibitemOpen
  \bibfield  {author} {\bibinfo {author} {\bibfnamefont {J.}~\bibnamefont
  {You}}, \bibinfo {author} {\bibfnamefont {S.}~\bibnamefont {Bongu}}, \bibinfo
  {author} {\bibfnamefont {Q.}~\bibnamefont {Bao}},\ and\ \bibinfo {author}
  {\bibfnamefont {N.}~\bibnamefont {Panoiu}},\ }\bibfield  {title} {\bibinfo
  {title} {Nonlinear optical properties and applications of 2d materials:
  theoretical and experimental aspects},\ }\href
  {https://doi.org/10.1515/nanoph-2018-0106} {\bibfield  {journal} {\bibinfo
  {journal} {Nanophotonics}\ }\textbf {\bibinfo {volume} {8}},\ \bibinfo
  {pages} {63} (\bibinfo {year} {2018})}\BibitemShut {NoStop}%
\bibitem [{\citenamefont {Raab}\ \emph {et~al.}(2019)\citenamefont {Raab},
  \citenamefont {Lange}, \citenamefont {Boland}, \citenamefont {Laepple},
  \citenamefont {Furthmeier}, \citenamefont {Dardanis}, \citenamefont
  {Dessmann}, \citenamefont {Li}, \citenamefont {Linfield}, \citenamefont
  {Davies}, \citenamefont {Vitiello},\ and\ \citenamefont {Huber}}]{Raab2019}%
  \BibitemOpen
  \bibfield  {author} {\bibinfo {author} {\bibfnamefont {J.}~\bibnamefont
  {Raab}}, \bibinfo {author} {\bibfnamefont {C.}~\bibnamefont {Lange}},
  \bibinfo {author} {\bibfnamefont {J.~L.}\ \bibnamefont {Boland}}, \bibinfo
  {author} {\bibfnamefont {I.}~\bibnamefont {Laepple}}, \bibinfo {author}
  {\bibfnamefont {M.}~\bibnamefont {Furthmeier}}, \bibinfo {author}
  {\bibfnamefont {E.}~\bibnamefont {Dardanis}}, \bibinfo {author}
  {\bibfnamefont {N.}~\bibnamefont {Dessmann}}, \bibinfo {author}
  {\bibfnamefont {L.}~\bibnamefont {Li}}, \bibinfo {author} {\bibfnamefont
  {E.~H.}\ \bibnamefont {Linfield}}, \bibinfo {author} {\bibfnamefont {A.~G.}\
  \bibnamefont {Davies}}, \bibinfo {author} {\bibfnamefont {M.~S.}\
  \bibnamefont {Vitiello}},\ and\ \bibinfo {author} {\bibfnamefont
  {R.}~\bibnamefont {Huber}},\ }\bibfield  {title} {\bibinfo {title} {Ultrafast
  two-dimensional field spectroscopy of terahertz intersubband saturable
  absorbers},\ }\href {https://doi.org/10.1364/oe.27.002248} {\bibfield
  {journal} {\bibinfo  {journal} {Opt. Express}\ }\textbf {\bibinfo {volume}
  {27}},\ \bibinfo {pages} {2248} (\bibinfo {year} {2019})}\BibitemShut
  {NoStop}%
\bibitem [{\citenamefont {Kovalev}\ \emph {et~al.}(2021)\citenamefont
  {Kovalev}, \citenamefont {Hafez}, \citenamefont {Tielrooij}, \citenamefont
  {Deinert}, \citenamefont {Ilyakov}, \citenamefont {Awari}, \citenamefont
  {Alcaraz}, \citenamefont {Soundarapandian}, \citenamefont {Saleta},
  \citenamefont {Germanskiy}, \citenamefont {Chen}, \citenamefont {Bawatna},
  \citenamefont {Green}, \citenamefont {Koppens}, \citenamefont {Mittendorff},
  \citenamefont {Bonn}, \citenamefont {Gensch},\ and\ \citenamefont
  {Turchinovich}}]{Kovalev2021}%
  \BibitemOpen
  \bibfield  {author} {\bibinfo {author} {\bibfnamefont {S.}~\bibnamefont
  {Kovalev}}, \bibinfo {author} {\bibfnamefont {H.~A.}\ \bibnamefont {Hafez}},
  \bibinfo {author} {\bibfnamefont {K.-J.}\ \bibnamefont {Tielrooij}}, \bibinfo
  {author} {\bibfnamefont {J.-C.}\ \bibnamefont {Deinert}}, \bibinfo {author}
  {\bibfnamefont {I.}~\bibnamefont {Ilyakov}}, \bibinfo {author} {\bibfnamefont
  {N.}~\bibnamefont {Awari}}, \bibinfo {author} {\bibfnamefont
  {D.}~\bibnamefont {Alcaraz}}, \bibinfo {author} {\bibfnamefont
  {K.}~\bibnamefont {Soundarapandian}}, \bibinfo {author} {\bibfnamefont
  {D.}~\bibnamefont {Saleta}}, \bibinfo {author} {\bibfnamefont
  {S.}~\bibnamefont {Germanskiy}}, \bibinfo {author} {\bibfnamefont
  {M.}~\bibnamefont {Chen}}, \bibinfo {author} {\bibfnamefont {M.}~\bibnamefont
  {Bawatna}}, \bibinfo {author} {\bibfnamefont {B.}~\bibnamefont {Green}},
  \bibinfo {author} {\bibfnamefont {F.~H.~L.}\ \bibnamefont {Koppens}},
  \bibinfo {author} {\bibfnamefont {M.}~\bibnamefont {Mittendorff}}, \bibinfo
  {author} {\bibfnamefont {M.}~\bibnamefont {Bonn}}, \bibinfo {author}
  {\bibfnamefont {M.}~\bibnamefont {Gensch}},\ and\ \bibinfo {author}
  {\bibfnamefont {D.}~\bibnamefont {Turchinovich}},\ }\bibfield  {title}
  {\bibinfo {title} {Electrical tunability of terahertz nonlinearity in
  graphene},\ }\href {https://doi.org/10.1126/sciadv.abf9809} {\bibfield
  {journal} {\bibinfo  {journal} {Sci. Adv.}\ }\textbf {\bibinfo {volume}
  {7}},\ \bibinfo {pages} {eabf9809} (\bibinfo {year} {2021})}\BibitemShut
  {NoStop}%
\bibitem [{\citenamefont {Mittendorff}\ \emph {et~al.}(2020)\citenamefont
  {Mittendorff}, \citenamefont {Winnerl},\ and\ \citenamefont
  {Murphy}}]{Mittendorff2021}%
  \BibitemOpen
  \bibfield  {author} {\bibinfo {author} {\bibfnamefont {M.}~\bibnamefont
  {Mittendorff}}, \bibinfo {author} {\bibfnamefont {S.}~\bibnamefont
  {Winnerl}},\ and\ \bibinfo {author} {\bibfnamefont {T.~E.}\ \bibnamefont
  {Murphy}},\ }\bibfield  {title} {\bibinfo {title} {2d {THz}
  optoelectronics},\ }\href {https://doi.org/10.1002/adom.202001500} {\bibfield
   {journal} {\bibinfo  {journal} {Adv. Opt. Mater.}\ }\textbf {\bibinfo
  {volume} {9}},\ \bibinfo {pages} {2001500} (\bibinfo {year}
  {2020})}\BibitemShut {NoStop}%
\bibitem [{\citenamefont {Dean}\ \emph {et~al.}(2010)\citenamefont {Dean},
  \citenamefont {Young}, \citenamefont {Meric}, \citenamefont {Lee},
  \citenamefont {Wang}, \citenamefont {Sorgenfrei}, \citenamefont {Watanabe},
  \citenamefont {Taniguchi}, \citenamefont {Kim}, \citenamefont {Shepard},\
  and\ \citenamefont {Hone}}]{Dean2010}%
  \BibitemOpen
  \bibfield  {author} {\bibinfo {author} {\bibfnamefont {C.~R.}\ \bibnamefont
  {Dean}}, \bibinfo {author} {\bibfnamefont {A.~F.}\ \bibnamefont {Young}},
  \bibinfo {author} {\bibfnamefont {I.}~\bibnamefont {Meric}}, \bibinfo
  {author} {\bibfnamefont {C.}~\bibnamefont {Lee}}, \bibinfo {author}
  {\bibfnamefont {L.}~\bibnamefont {Wang}}, \bibinfo {author} {\bibfnamefont
  {S.}~\bibnamefont {Sorgenfrei}}, \bibinfo {author} {\bibfnamefont
  {K.}~\bibnamefont {Watanabe}}, \bibinfo {author} {\bibfnamefont
  {T.}~\bibnamefont {Taniguchi}}, \bibinfo {author} {\bibfnamefont
  {P.}~\bibnamefont {Kim}}, \bibinfo {author} {\bibfnamefont {K.~L.}\
  \bibnamefont {Shepard}},\ and\ \bibinfo {author} {\bibfnamefont
  {J.}~\bibnamefont {Hone}},\ }\bibfield  {title} {\bibinfo {title} {Boron
  nitride substrates for high-quality graphene electronics},\ }\href
  {https://doi.org/10.1038/nnano.2010.172} {\bibfield  {journal} {\bibinfo
  {journal} {Nat. Nanotechnol.}\ }\textbf {\bibinfo {volume} {5}},\ \bibinfo
  {pages} {722} (\bibinfo {year} {2010})}\BibitemShut {NoStop}%
\bibitem [{\citenamefont {Wang}\ \emph {et~al.}(2013)\citenamefont {Wang},
  \citenamefont {Meric}, \citenamefont {Huang}, \citenamefont {Gao},
  \citenamefont {Gao}, \citenamefont {Tran}, \citenamefont {Taniguchi},
  \citenamefont {Watanabe}, \citenamefont {Campos}, \citenamefont {Muller},
  \citenamefont {Guo}, \citenamefont {Kim}, \citenamefont {Hone}, \citenamefont
  {Shepard},\ and\ \citenamefont {Dean}}]{Wang2013}%
  \BibitemOpen
  \bibfield  {author} {\bibinfo {author} {\bibfnamefont {L.}~\bibnamefont
  {Wang}}, \bibinfo {author} {\bibfnamefont {I.}~\bibnamefont {Meric}},
  \bibinfo {author} {\bibfnamefont {P.~Y.}\ \bibnamefont {Huang}}, \bibinfo
  {author} {\bibfnamefont {Q.}~\bibnamefont {Gao}}, \bibinfo {author}
  {\bibfnamefont {Y.}~\bibnamefont {Gao}}, \bibinfo {author} {\bibfnamefont
  {H.}~\bibnamefont {Tran}}, \bibinfo {author} {\bibfnamefont {T.}~\bibnamefont
  {Taniguchi}}, \bibinfo {author} {\bibfnamefont {K.}~\bibnamefont {Watanabe}},
  \bibinfo {author} {\bibfnamefont {L.~M.}\ \bibnamefont {Campos}}, \bibinfo
  {author} {\bibfnamefont {D.~A.}\ \bibnamefont {Muller}}, \bibinfo {author}
  {\bibfnamefont {J.}~\bibnamefont {Guo}}, \bibinfo {author} {\bibfnamefont
  {P.}~\bibnamefont {Kim}}, \bibinfo {author} {\bibfnamefont {J.}~\bibnamefont
  {Hone}}, \bibinfo {author} {\bibfnamefont {K.~L.}\ \bibnamefont {Shepard}},\
  and\ \bibinfo {author} {\bibfnamefont {C.~R.}\ \bibnamefont {Dean}},\
  }\bibfield  {title} {\bibinfo {title} {One-dimensional electrical contact to
  a two-dimensional material},\ }\href
  {https://doi.org/10.1126/science.1244358} {\bibfield  {journal} {\bibinfo
  {journal} {Science}\ }\textbf {\bibinfo {volume} {342}},\ \bibinfo {pages}
  {614} (\bibinfo {year} {2013})}\BibitemShut {NoStop}%
\bibitem [{\citenamefont {Sandner}\ \emph {et~al.}(2015)\citenamefont
  {Sandner}, \citenamefont {Preis}, \citenamefont {Schell}, \citenamefont
  {Giudici}, \citenamefont {Watanabe}, \citenamefont {Taniguchi}, \citenamefont
  {Weiss},\ and\ \citenamefont {Eroms}}]{Sandner2015}%
  \BibitemOpen
  \bibfield  {author} {\bibinfo {author} {\bibfnamefont {A.}~\bibnamefont
  {Sandner}}, \bibinfo {author} {\bibfnamefont {T.}~\bibnamefont {Preis}},
  \bibinfo {author} {\bibfnamefont {C.}~\bibnamefont {Schell}}, \bibinfo
  {author} {\bibfnamefont {P.}~\bibnamefont {Giudici}}, \bibinfo {author}
  {\bibfnamefont {K.}~\bibnamefont {Watanabe}}, \bibinfo {author}
  {\bibfnamefont {T.}~\bibnamefont {Taniguchi}}, \bibinfo {author}
  {\bibfnamefont {D.}~\bibnamefont {Weiss}},\ and\ \bibinfo {author}
  {\bibfnamefont {J.}~\bibnamefont {Eroms}},\ }\bibfield  {title} {\bibinfo
  {title} {Ballistic transport in graphene antidot lattices},\ }\href
  {https://doi.org/10.1021/acs.nanolett.5b04414} {\bibfield  {journal}
  {\bibinfo  {journal} {Nano Lett.}\ }\textbf {\bibinfo {volume} {15}},\
  \bibinfo {pages} {8402} (\bibinfo {year} {2015})}\BibitemShut {NoStop}%
\bibitem [{\citenamefont {Ganichev}\ \emph {et~al.}(1982)\citenamefont
  {Ganichev}, \citenamefont {Emel'yanov},\ and\ \citenamefont
  {Yaroshetskii}}]{Ganichev1982}%
  \BibitemOpen
  \bibfield  {author} {\bibinfo {author} {\bibfnamefont {S.~D.}\ \bibnamefont
  {Ganichev}}, \bibinfo {author} {\bibfnamefont {S.~A.}\ \bibnamefont
  {Emel'yanov}},\ and\ \bibinfo {author} {\bibfnamefont {I.~D.}\ \bibnamefont
  {Yaroshetskii}},\ }\href@noop {} {\bibfield  {journal} {\bibinfo  {journal}
  {JETP Lett.}\ }\textbf {\bibinfo {volume} {35}},\ \bibinfo {pages} {368}
  (\bibinfo {year} {1982})}\BibitemShut {NoStop}%
\bibitem [{\citenamefont {Shalygin}\ \emph {et~al.}(2007)\citenamefont
  {Shalygin}, \citenamefont {Diehl}, \citenamefont {Hoffmann}, \citenamefont
  {Danilov}, \citenamefont {Herrle}, \citenamefont {Tarasenko}, \citenamefont
  {Schuh}, \citenamefont {Gerl}, \citenamefont {Wegscheider}, \citenamefont
  {Prettl},\ and\ \citenamefont {Ganichev}}]{Shalygin2006}%
  \BibitemOpen
  \bibfield  {author} {\bibinfo {author} {\bibfnamefont {V.~A.}\ \bibnamefont
  {Shalygin}}, \bibinfo {author} {\bibfnamefont {H.}~\bibnamefont {Diehl}},
  \bibinfo {author} {\bibfnamefont {C.}~\bibnamefont {Hoffmann}}, \bibinfo
  {author} {\bibfnamefont {S.~N.}\ \bibnamefont {Danilov}}, \bibinfo {author}
  {\bibfnamefont {T.}~\bibnamefont {Herrle}}, \bibinfo {author} {\bibfnamefont
  {S.~A.}\ \bibnamefont {Tarasenko}}, \bibinfo {author} {\bibfnamefont
  {D.}~\bibnamefont {Schuh}}, \bibinfo {author} {\bibfnamefont
  {C.}~\bibnamefont {Gerl}}, \bibinfo {author} {\bibfnamefont {W.}~\bibnamefont
  {Wegscheider}}, \bibinfo {author} {\bibfnamefont {W.}~\bibnamefont
  {Prettl}},\ and\ \bibinfo {author} {\bibfnamefont {S.~D.}\ \bibnamefont
  {Ganichev}},\ }\bibfield  {title} {\bibinfo {title} {Spin photocurrents and
  the circular photon drag effect in (110)-grown quantum well structures},\
  }\href {https://doi.org/10.1134/s0021364006220097} {\bibfield  {journal}
  {\bibinfo  {journal} {{JETP} Letters}\ }\textbf {\bibinfo {volume} {84}},\
  \bibinfo {pages} {570} (\bibinfo {year} {2007})}\BibitemShut {NoStop}%
\bibitem [{\citenamefont {Plank}\ \emph {et~al.}(2016)\citenamefont {Plank},
  \citenamefont {Golub}, \citenamefont {Bauer}, \citenamefont {Bel'kov},
  \citenamefont {Herrmann}, \citenamefont {Olbrich}, \citenamefont {Eschbach},
  \citenamefont {Plucinski}, \citenamefont {Schneider}, \citenamefont
  {Kampmeier}, \citenamefont {Lanius}, \citenamefont {Mussler}, \citenamefont
  {Grützmacher},\ and\ \citenamefont {Ganichev}}]{Plank2016drag}%
  \BibitemOpen
  \bibfield  {author} {\bibinfo {author} {\bibfnamefont {H.}~\bibnamefont
  {Plank}}, \bibinfo {author} {\bibfnamefont {L.~E.}\ \bibnamefont {Golub}},
  \bibinfo {author} {\bibfnamefont {S.}~\bibnamefont {Bauer}}, \bibinfo
  {author} {\bibfnamefont {V.~V.}\ \bibnamefont {Bel'kov}}, \bibinfo {author}
  {\bibfnamefont {T.}~\bibnamefont {Herrmann}}, \bibinfo {author}
  {\bibfnamefont {P.}~\bibnamefont {Olbrich}}, \bibinfo {author} {\bibfnamefont
  {M.}~\bibnamefont {Eschbach}}, \bibinfo {author} {\bibfnamefont
  {L.}~\bibnamefont {Plucinski}}, \bibinfo {author} {\bibfnamefont {C.~M.}\
  \bibnamefont {Schneider}}, \bibinfo {author} {\bibfnamefont {J.}~\bibnamefont
  {Kampmeier}}, \bibinfo {author} {\bibfnamefont {M.}~\bibnamefont {Lanius}},
  \bibinfo {author} {\bibfnamefont {G.}~\bibnamefont {Mussler}}, \bibinfo
  {author} {\bibfnamefont {D.}~\bibnamefont {Grützmacher}},\ and\ \bibinfo
  {author} {\bibfnamefont {S.~D.}\ \bibnamefont {Ganichev}},\ }\bibfield
  {title} {\bibinfo {title} {Photon drag effect in (bi$_{1-x}$sb$_x$)$_2$te$_3$
  three-dimensional topological insulators},\ }\href
  {https://doi.org/10.1103/physrevb.93.125434} {\bibfield  {journal} {\bibinfo
  {journal} {Phys. Rev. B}\ }\textbf {\bibinfo {volume} {93}},\ \bibinfo
  {pages} {125434} (\bibinfo {year} {2016})}\BibitemShut {NoStop}%
\bibitem [{\citenamefont {Ganichev}\ \emph {et~al.}(2003)\citenamefont
  {Ganichev}, \citenamefont {Schneider}, \citenamefont {Bel'kov}, \citenamefont
  {Ivchenko}, \citenamefont {Tarasenko}, \citenamefont {Wegscheider},
  \citenamefont {Weiss}, \citenamefont {Schuh}, \citenamefont {Murdin},
  \citenamefont {Phillips}, \citenamefont {Pidgeon}, \citenamefont {Clarke},
  \citenamefont {Merrick}, \citenamefont {Murzyn}, \citenamefont {Beregulin},\
  and\ \citenamefont {Prettl}}]{Ganichev2003}%
  \BibitemOpen
  \bibfield  {author} {\bibinfo {author} {\bibfnamefont {S.~D.}\ \bibnamefont
  {Ganichev}}, \bibinfo {author} {\bibfnamefont {P.}~\bibnamefont {Schneider}},
  \bibinfo {author} {\bibfnamefont {V.~V.}\ \bibnamefont {Bel'kov}}, \bibinfo
  {author} {\bibfnamefont {E.~L.}\ \bibnamefont {Ivchenko}}, \bibinfo {author}
  {\bibfnamefont {S.~A.}\ \bibnamefont {Tarasenko}}, \bibinfo {author}
  {\bibfnamefont {W.}~\bibnamefont {Wegscheider}}, \bibinfo {author}
  {\bibfnamefont {D.}~\bibnamefont {Weiss}}, \bibinfo {author} {\bibfnamefont
  {D.}~\bibnamefont {Schuh}}, \bibinfo {author} {\bibfnamefont {B.~N.}\
  \bibnamefont {Murdin}}, \bibinfo {author} {\bibfnamefont {P.~J.}\
  \bibnamefont {Phillips}}, \bibinfo {author} {\bibfnamefont {C.~R.}\
  \bibnamefont {Pidgeon}}, \bibinfo {author} {\bibfnamefont {D.~G.}\
  \bibnamefont {Clarke}}, \bibinfo {author} {\bibfnamefont {M.}~\bibnamefont
  {Merrick}}, \bibinfo {author} {\bibfnamefont {P.}~\bibnamefont {Murzyn}},
  \bibinfo {author} {\bibfnamefont {E.~V.}\ \bibnamefont {Beregulin}},\ and\
  \bibinfo {author} {\bibfnamefont {W.}~\bibnamefont {Prettl}},\ }\bibfield
  {title} {\bibinfo {title} {Spin-galvanic effect due to optical spin
  orientation in n-type {GaAs} quantum well structures},\ }\href
  {https://doi.org/10.1103/physrevb.68.081302} {\bibfield  {journal} {\bibinfo
  {journal} {Phys. Rev. B}\ }\textbf {\bibinfo {volume} {68}},\ \bibinfo
  {pages} {081302(R)} (\bibinfo {year} {2003})}\BibitemShut {NoStop}%
\bibitem [{\citenamefont {Ganichev}\ \emph {et~al.}(1989)\citenamefont
  {Ganichev}, \citenamefont {Terent'ev},\ and\ \citenamefont
  {Yaroshetskii}}]{Ganichev84p20}%
  \BibitemOpen
  \bibfield  {author} {\bibinfo {author} {\bibfnamefont {S.~D.}\ \bibnamefont
  {Ganichev}}, \bibinfo {author} {\bibfnamefont {Y.~V.}\ \bibnamefont
  {Terent'ev}},\ and\ \bibinfo {author} {\bibfnamefont {I.~D.}\ \bibnamefont
  {Yaroshetskii}},\ }\bibfield  {title} {\bibinfo {title} {Photon-drag
  photodetectors for the far-ir and submillimeter regions},\ }\href@noop {}
  {\bibfield  {journal} {\bibinfo  {journal} {Sov. Tech. Phys. Lett.}\ }\textbf
  {\bibinfo {volume} {11}},\ \bibinfo {pages} {20} (\bibinfo {year}
  {1989})}\BibitemShut {NoStop}%
\bibitem [{\citenamefont {Castilla}\ \emph {et~al.}(2019)\citenamefont
  {Castilla}, \citenamefont {Terr{\'{e}}s}, \citenamefont {Autore},
  \citenamefont {Viti}, \citenamefont {Li}, \citenamefont {Nikitin},
  \citenamefont {Vangelidis}, \citenamefont {Watanabe}, \citenamefont
  {Taniguchi}, \citenamefont {Lidorikis}, \citenamefont {Vitiello},
  \citenamefont {Hillenbrand}, \citenamefont {Tielrooij},\ and\ \citenamefont
  {Koppens}}]{Castilla}%
  \BibitemOpen
  \bibfield  {author} {\bibinfo {author} {\bibfnamefont {S.}~\bibnamefont
  {Castilla}}, \bibinfo {author} {\bibfnamefont {B.}~\bibnamefont
  {Terr{\'{e}}s}}, \bibinfo {author} {\bibfnamefont {M.}~\bibnamefont
  {Autore}}, \bibinfo {author} {\bibfnamefont {L.}~\bibnamefont {Viti}},
  \bibinfo {author} {\bibfnamefont {J.}~\bibnamefont {Li}}, \bibinfo {author}
  {\bibfnamefont {A.~Y.}\ \bibnamefont {Nikitin}}, \bibinfo {author}
  {\bibfnamefont {I.}~\bibnamefont {Vangelidis}}, \bibinfo {author}
  {\bibfnamefont {K.}~\bibnamefont {Watanabe}}, \bibinfo {author}
  {\bibfnamefont {T.}~\bibnamefont {Taniguchi}}, \bibinfo {author}
  {\bibfnamefont {E.}~\bibnamefont {Lidorikis}}, \bibinfo {author}
  {\bibfnamefont {M.~S.}\ \bibnamefont {Vitiello}}, \bibinfo {author}
  {\bibfnamefont {R.}~\bibnamefont {Hillenbrand}}, \bibinfo {author}
  {\bibfnamefont {K.-J.}\ \bibnamefont {Tielrooij}},\ and\ \bibinfo {author}
  {\bibfnamefont {F.~H.}\ \bibnamefont {Koppens}},\ }\bibfield  {title}
  {\bibinfo {title} {Fast and sensitive terahertz detection using an
  antenna-integrated graphene pn junction},\ }\href
  {https://doi.org/10.1021/acs.nanolett.8b04171} {\bibfield  {journal}
  {\bibinfo  {journal} {Nano Lett.}\ }\textbf {\bibinfo {volume} {19}},\
  \bibinfo {pages} {2765} (\bibinfo {year} {2019})}\BibitemShut {NoStop}%
\bibitem [{\citenamefont {Ganichev}\ and\ \citenamefont
  {Prettl}(2006)}]{Ganichevbook}%
  \BibitemOpen
  \bibfield  {author} {\bibinfo {author} {\bibfnamefont {S.~D.}\ \bibnamefont
  {Ganichev}}\ and\ \bibinfo {author} {\bibfnamefont {W.}~\bibnamefont
  {Prettl}},\ }\href@noop {} {\emph {\bibinfo {title} {Intense Terahertz
  Excitation of Semiconductors}}}\ (\bibinfo  {publisher} {Oxford University
  Press 2006},\ \bibinfo {year} {2006})\BibitemShut {NoStop}%
\bibitem [{\citenamefont {Sarkar}\ \emph {et~al.}(2015)\citenamefont {Sarkar},
  \citenamefont {Amin}, \citenamefont {Modak}, \citenamefont {Singh},
  \citenamefont {Mukerjee},\ and\ \citenamefont {Bid}}]{Sarkar2015}%
  \BibitemOpen
  \bibfield  {author} {\bibinfo {author} {\bibfnamefont {S.}~\bibnamefont
  {Sarkar}}, \bibinfo {author} {\bibfnamefont {K.~R.}\ \bibnamefont {Amin}},
  \bibinfo {author} {\bibfnamefont {R.}~\bibnamefont {Modak}}, \bibinfo
  {author} {\bibfnamefont {A.}~\bibnamefont {Singh}}, \bibinfo {author}
  {\bibfnamefont {S.}~\bibnamefont {Mukerjee}},\ and\ \bibinfo {author}
  {\bibfnamefont {A.}~\bibnamefont {Bid}},\ }\bibfield  {title} {\bibinfo
  {title} {Role of different scattering mechanisms on the temperature
  dependence of transport in graphene},\ }\href
  {https://doi.org/10.1038/srep16772} {\bibfield  {journal} {\bibinfo
  {journal} {Sci. Rep.}\ }\textbf {\bibinfo {volume} {5}},\ \bibinfo {pages}
  {16772} (\bibinfo {year} {2015})}\BibitemShut {NoStop}%
\bibitem [{\citenamefont {Tarasenko}(2011)}]{Tarasenko2011}%
  \BibitemOpen
  \bibfield  {author} {\bibinfo {author} {\bibfnamefont {S.~A.}\ \bibnamefont
  {Tarasenko}},\ }\bibfield  {title} {\bibinfo {title} {Direct current driven
  by ac electric field in quantum wells},\ }\href
  {https://doi.org/10.1103/physrevb.83.035313} {\bibfield  {journal} {\bibinfo
  {journal} {Phys. Rev. B}\ }\textbf {\bibinfo {volume} {83}},\ \bibinfo
  {pages} {035313} (\bibinfo {year} {2011})}\BibitemShut {NoStop}%
\bibitem [{\citenamefont {Golub}\ \emph {et~al.}(2011)\citenamefont {Golub},
  \citenamefont {Tarasenko}, \citenamefont {Entin},\ and\ \citenamefont
  {Magarill}}]{Golub2011}%
  \BibitemOpen
  \bibfield  {author} {\bibinfo {author} {\bibfnamefont {L.~E.}\ \bibnamefont
  {Golub}}, \bibinfo {author} {\bibfnamefont {S.~A.}\ \bibnamefont
  {Tarasenko}}, \bibinfo {author} {\bibfnamefont {M.~V.}\ \bibnamefont
  {Entin}},\ and\ \bibinfo {author} {\bibfnamefont {L.~I.}\ \bibnamefont
  {Magarill}},\ }\bibfield  {title} {\bibinfo {title} {Valley separation in
  graphene by polarized light},\ }\href
  {https://doi.org/10.1103/physrevb.84.195408} {\bibfield  {journal} {\bibinfo
  {journal} {Phys. Rev. B}\ }\textbf {\bibinfo {volume} {84}},\ \bibinfo
  {pages} {195408} (\bibinfo {year} {2011})}\BibitemShut {NoStop}%
\bibitem [{\citenamefont {Hartmann}\ and\ \citenamefont
  {Portnoi}(2011)}]{Hartmann2011}%
  \BibitemOpen
  \bibfield  {author} {\bibinfo {author} {\bibfnamefont {R.~R.}\ \bibnamefont
  {Hartmann}}\ and\ \bibinfo {author} {\bibfnamefont {M.~E.}\ \bibnamefont
  {Portnoi}},\ }\href@noop {} {\emph {\bibinfo {title} {Optoelectronic
  Properties of Carbon-based Nanostructures: Steering electrons in graphene by
  electromagnetic fields}}}\ (\bibinfo  {publisher} {LAP LAMBERT Academic
  Publishing, Saarbrucken},\ \bibinfo {year} {2011})\BibitemShut {NoStop}%
\bibitem [{\citenamefont {Durnev}\ and\ \citenamefont
  {Tarasenko}()}]{DurnevTarasenko}%
  \BibitemOpen
  \bibfield  {author} {\bibinfo {author} {\bibfnamefont {M.~V.}\ \bibnamefont
  {Durnev}}\ and\ \bibinfo {author} {\bibfnamefont {S.~A.}\ \bibnamefont
  {Tarasenko}},\ }\href@noop {} {\bibinfo  {journal} {private communication}\
  }\BibitemShut {NoStop}%
\bibitem [{\citenamefont {Novikov}(2007)}]{Novikov2007}%
  \BibitemOpen
\bibfield  {journal} {  }\bibfield  {author} {\bibinfo {author} {\bibfnamefont
  {D.~S.}\ \bibnamefont {Novikov}},\ }\bibfield  {title} {\bibinfo {title}
  {Numbers of donors and acceptors from transport measurements in graphene},\
  }\href {https://doi.org/10.1063/1.2779107} {\bibfield  {journal} {\bibinfo
  {journal} {Appl. Phys. Lett.}\ }\textbf {\bibinfo {volume} {91}},\ \bibinfo
  {pages} {102102} (\bibinfo {year} {2007})}\BibitemShut {NoStop}%
\bibitem [{\citenamefont {Klatt}\ \emph {et~al.}(2010)\citenamefont {Klatt},
  \citenamefont {Hilser}, \citenamefont {Qiao}, \citenamefont {Beck},
  \citenamefont {Gebs}, \citenamefont {Bartels}, \citenamefont {Huska},
  \citenamefont {Lemmer}, \citenamefont {Bastian}, \citenamefont {Johnston},
  \citenamefont {Fischer}, \citenamefont {Faist},\ and\ \citenamefont
  {Dekorsy}}]{Klatt2010}%
  \BibitemOpen
  \bibfield  {author} {\bibinfo {author} {\bibfnamefont {G.}~\bibnamefont
  {Klatt}}, \bibinfo {author} {\bibfnamefont {F.}~\bibnamefont {Hilser}},
  \bibinfo {author} {\bibfnamefont {W.}~\bibnamefont {Qiao}}, \bibinfo {author}
  {\bibfnamefont {M.}~\bibnamefont {Beck}}, \bibinfo {author} {\bibfnamefont
  {R.}~\bibnamefont {Gebs}}, \bibinfo {author} {\bibfnamefont {A.}~\bibnamefont
  {Bartels}}, \bibinfo {author} {\bibfnamefont {K.}~\bibnamefont {Huska}},
  \bibinfo {author} {\bibfnamefont {U.}~\bibnamefont {Lemmer}}, \bibinfo
  {author} {\bibfnamefont {G.}~\bibnamefont {Bastian}}, \bibinfo {author}
  {\bibfnamefont {M.}~\bibnamefont {Johnston}}, \bibinfo {author}
  {\bibfnamefont {M.}~\bibnamefont {Fischer}}, \bibinfo {author} {\bibfnamefont
  {J.}~\bibnamefont {Faist}},\ and\ \bibinfo {author} {\bibfnamefont
  {T.}~\bibnamefont {Dekorsy}},\ }\bibfield  {title} {\bibinfo {title}
  {Terahertz emission from lateralphoto-dember currents},\ }\href
  {https://doi.org/10.1364/oe.18.004939} {\bibfield  {journal} {\bibinfo
  {journal} {Opt. Express}\ }\textbf {\bibinfo {volume} {18}},\ \bibinfo
  {pages} {4939} (\bibinfo {year} {2010})}\BibitemShut {NoStop}%
\bibitem [{\citenamefont {Obraztsov}\ \emph {et~al.}(2014)\citenamefont
  {Obraztsov}, \citenamefont {Kanda}, \citenamefont {Konishi}, \citenamefont
  {Kuwata-Gonokami}, \citenamefont {Garnov}, \citenamefont {Obraztsov},\ and\
  \citenamefont {Svirko}}]{Obraztsov2014}%
  \BibitemOpen
  \bibfield  {author} {\bibinfo {author} {\bibfnamefont {P.~A.}\ \bibnamefont
  {Obraztsov}}, \bibinfo {author} {\bibfnamefont {N.}~\bibnamefont {Kanda}},
  \bibinfo {author} {\bibfnamefont {K.}~\bibnamefont {Konishi}}, \bibinfo
  {author} {\bibfnamefont {M.}~\bibnamefont {Kuwata-Gonokami}}, \bibinfo
  {author} {\bibfnamefont {S.~V.}\ \bibnamefont {Garnov}}, \bibinfo {author}
  {\bibfnamefont {A.~N.}\ \bibnamefont {Obraztsov}},\ and\ \bibinfo {author}
  {\bibfnamefont {Y.~P.}\ \bibnamefont {Svirko}},\ }\bibfield  {title}
  {\bibinfo {title} {Photon-drag-induced terahertz emission from graphene},\
  }\href {https://doi.org/10.1103/physrevb.90.241416} {\bibfield  {journal}
  {\bibinfo  {journal} {Phys. Rev. B}\ }\textbf {\bibinfo {volume} {90}},\
  \bibinfo {pages} {241416(R)} (\bibinfo {year} {2014})}\BibitemShut {NoStop}%
\bibitem [{\citenamefont {Li}\ \emph {et~al.}(2017)\citenamefont {Li},
  \citenamefont {Lin}, \citenamefont {Rui}, \citenamefont {Li}, \citenamefont
  {Zhang}, \citenamefont {Kang}, \citenamefont {Zhang}, \citenamefont {Peng},
  \citenamefont {Liu},\ and\ \citenamefont {Xu}}]{Li2017}%
  \BibitemOpen
  \bibfield  {author} {\bibinfo {author} {\bibfnamefont {J.}~\bibnamefont
  {Li}}, \bibinfo {author} {\bibfnamefont {L.}~\bibnamefont {Lin}}, \bibinfo
  {author} {\bibfnamefont {D.}~\bibnamefont {Rui}}, \bibinfo {author}
  {\bibfnamefont {Q.}~\bibnamefont {Li}}, \bibinfo {author} {\bibfnamefont
  {J.}~\bibnamefont {Zhang}}, \bibinfo {author} {\bibfnamefont
  {N.}~\bibnamefont {Kang}}, \bibinfo {author} {\bibfnamefont {Y.}~\bibnamefont
  {Zhang}}, \bibinfo {author} {\bibfnamefont {H.}~\bibnamefont {Peng}},
  \bibinfo {author} {\bibfnamefont {Z.}~\bibnamefont {Liu}},\ and\ \bibinfo
  {author} {\bibfnamefont {H.~Q.}\ \bibnamefont {Xu}},\ }\bibfield  {title}
  {\bibinfo {title} {Electron{\textendash}hole symmetry breaking in charge
  transport in nitrogen-doped graphene},\ }\href
  {https://doi.org/10.1021/acsnano.7b00313} {\bibfield  {journal} {\bibinfo
  {journal} {{ACS} Nano}\ }\textbf {\bibinfo {volume} {11}},\ \bibinfo {pages}
  {4641} (\bibinfo {year} {2017})}\BibitemShut {NoStop}%
\bibitem [{\citenamefont {Singh}\ \emph {et~al.}(2017)\citenamefont {Singh},
  \citenamefont {Bolotin}, \citenamefont {Ghosh},\ and\ \citenamefont
  {Agarwal}}]{Nonlin_abs_Agarwal}%
  \BibitemOpen
  \bibfield  {author} {\bibinfo {author} {\bibfnamefont {A.}~\bibnamefont
  {Singh}}, \bibinfo {author} {\bibfnamefont {K.~I.}\ \bibnamefont {Bolotin}},
  \bibinfo {author} {\bibfnamefont {S.}~\bibnamefont {Ghosh}},\ and\ \bibinfo
  {author} {\bibfnamefont {A.}~\bibnamefont {Agarwal}},\ }\bibfield  {title}
  {\bibinfo {title} {Nonlinear optical conductivity of a generic two-band
  system with application to doped and gapped graphene},\ }\href
  {https://doi.org/10.1103/PhysRevB.95.155421} {\bibfield  {journal} {\bibinfo
  {journal} {Phys. Rev. B}\ }\textbf {\bibinfo {volume} {95}},\ \bibinfo
  {pages} {155421} (\bibinfo {year} {2017})}\BibitemShut {NoStop}%
\bibitem [{\citenamefont {Leppenen}\ \emph {et~al.}(2019)\citenamefont
  {Leppenen}, \citenamefont {Ivchenko},\ and\ \citenamefont
  {Golub}}]{pssb_2019}%
  \BibitemOpen
  \bibfield  {author} {\bibinfo {author} {\bibfnamefont {N.~V.}\ \bibnamefont
  {Leppenen}}, \bibinfo {author} {\bibfnamefont {E.~L.}\ \bibnamefont
  {Ivchenko}},\ and\ \bibinfo {author} {\bibfnamefont {L.~E.}\ \bibnamefont
  {Golub}},\ }\bibfield  {title} {\bibinfo {title} {Nonlinear absorption and
  photocurrent in weyl semimetals},\ }\href
  {https://doi.org/10.1002/pssb.201900305} {\bibfield  {journal} {\bibinfo
  {journal} {physica status solidi (b)}\ }\textbf {\bibinfo {volume} {256}},\
  \bibinfo {pages} {1900305} (\bibinfo {year} {2019})}\BibitemShut {NoStop}%
\bibitem [{\citenamefont {Matsyshyn}\ \emph {et~al.}(2021)\citenamefont
  {Matsyshyn}, \citenamefont {Piazza}, \citenamefont {Moessner},\ and\
  \citenamefont {Sodemann}}]{Matsyshyn2021}%
  \BibitemOpen
  \bibfield  {author} {\bibinfo {author} {\bibfnamefont {O.}~\bibnamefont
  {Matsyshyn}}, \bibinfo {author} {\bibfnamefont {F.}~\bibnamefont {Piazza}},
  \bibinfo {author} {\bibfnamefont {R.}~\bibnamefont {Moessner}},\ and\
  \bibinfo {author} {\bibfnamefont {I.}~\bibnamefont {Sodemann}},\ }\bibfield
  {title} {\bibinfo {title} {The rabi regime of current rectification in
  solids},\ }\href@noop {} {\bibfield  {journal} {\bibinfo  {journal} {arXiv}\
  } (\bibinfo {year} {2021})},\ \Eprint {https://arxiv.org/abs/2104.00689}
  {2104.00689} \BibitemShut {NoStop}%
\bibitem [{\citenamefont {S.D.Ganichev}\ \emph {et~al.}(1983)\citenamefont
  {S.D.Ganichev}, \citenamefont {Emel'yanov}, \citenamefont {Ivchenko},
  \citenamefont {Perlin},\ and\ \citenamefont
  {I.D.Yaroshetskii}}]{Ganichev1983}%
  \BibitemOpen
  \bibfield  {author} {\bibinfo {author} {\bibnamefont {S.D.Ganichev}},
  \bibinfo {author} {\bibfnamefont {S.}~\bibnamefont {Emel'yanov}}, \bibinfo
  {author} {\bibfnamefont {E.}~\bibnamefont {Ivchenko}}, \bibinfo {author}
  {\bibfnamefont {E.}~\bibnamefont {Perlin}},\ and\ \bibinfo {author}
  {\bibnamefont {I.D.Yaroshetskii}},\ }\bibfield  {title} {\bibinfo {title}
  {Many-photon absorption in p-ge in the submillimeter range},\ }\href@noop {}
  {\bibfield  {journal} {\bibinfo  {journal} {Sov. Phys. JETP Lett.}\ }\textbf
  {\bibinfo {volume} {37}},\ \bibinfo {pages} {568} (\bibinfo {year}
  {1983})}\BibitemShut {NoStop}%
\bibitem [{\citenamefont {Ganichev}\ \emph {et~al.}(1993)\citenamefont
  {Ganichev}, \citenamefont {Ivchenko}, \citenamefont {Rasulov}, \citenamefont
  {Yaroshetskii},\ and\ \citenamefont {Averbukh}}]{Ganichev1993}%
  \BibitemOpen
  \bibfield  {author} {\bibinfo {author} {\bibfnamefont {S.}~\bibnamefont
  {Ganichev}}, \bibinfo {author} {\bibfnamefont {E.}~\bibnamefont {Ivchenko}},
  \bibinfo {author} {\bibfnamefont {R.}~\bibnamefont {Rasulov}}, \bibinfo
  {author} {\bibfnamefont {I.}~\bibnamefont {Yaroshetskii}},\ and\ \bibinfo
  {author} {\bibfnamefont {B.}~\bibnamefont {Averbukh}},\ }\bibfield  {title}
  {\bibinfo {title} {Linear-circular dichroism of photon drag effect at
  nonlinear intersubband absorption of light in p-type ge},\ }\href
  {https://doi.org/10.5283/EPUB.2261} {\bibfield  {journal} {\bibinfo
  {journal} {Phys. Solid State}\ }\textbf {\bibinfo {volume} {35}},\ \bibinfo
  {pages} {104} (\bibinfo {year} {1993})}\BibitemShut {NoStop}%
\bibitem [{\citenamefont {Yang}\ \emph {et~al.}(2011)\citenamefont {Yang},
  \citenamefont {Feng}, \citenamefont {Wang}, \citenamefont {Huang},
  \citenamefont {Chen}, \citenamefont {Wee},\ and\ \citenamefont
  {Ji}}]{Yang2011}%
  \BibitemOpen
  \bibfield  {author} {\bibinfo {author} {\bibfnamefont {H.}~\bibnamefont
  {Yang}}, \bibinfo {author} {\bibfnamefont {X.}~\bibnamefont {Feng}}, \bibinfo
  {author} {\bibfnamefont {Q.}~\bibnamefont {Wang}}, \bibinfo {author}
  {\bibfnamefont {H.}~\bibnamefont {Huang}}, \bibinfo {author} {\bibfnamefont
  {W.}~\bibnamefont {Chen}}, \bibinfo {author} {\bibfnamefont {A.~T.~S.}\
  \bibnamefont {Wee}},\ and\ \bibinfo {author} {\bibfnamefont {W.}~\bibnamefont
  {Ji}},\ }\bibfield  {title} {\bibinfo {title} {Giant two-photon absorption in
  bilayer graphene},\ }\href {https://doi.org/10.1021/nl200587h} {\bibfield
  {journal} {\bibinfo  {journal} {Nano Lett.}\ }\textbf {\bibinfo {volume}
  {11}},\ \bibinfo {pages} {2622} (\bibinfo {year} {2011})}\BibitemShut
  {NoStop}%
\bibitem [{\citenamefont {Rioux}\ \emph {et~al.}(2011)\citenamefont {Rioux},
  \citenamefont {Burkard},\ and\ \citenamefont {Sipe}}]{Rioux2011}%
  \BibitemOpen
  \bibfield  {author} {\bibinfo {author} {\bibfnamefont {J.}~\bibnamefont
  {Rioux}}, \bibinfo {author} {\bibfnamefont {G.}~\bibnamefont {Burkard}},\
  and\ \bibinfo {author} {\bibfnamefont {J.~E.}\ \bibnamefont {Sipe}},\
  }\bibfield  {title} {\bibinfo {title} {Current injection by coherent one- and
  two-photon excitation in graphene and its bilayer},\ }\href
  {https://doi.org/10.1103/physrevb.83.195406} {\bibfield  {journal} {\bibinfo
  {journal} {Phys. Rev. B}\ }\textbf {\bibinfo {volume} {83}},\ \bibinfo
  {pages} {195406} (\bibinfo {year} {2011})}\BibitemShut {NoStop}%
\bibitem [{\citenamefont {Brown}\ \emph {et~al.}(2016)\citenamefont {Brown},
  \citenamefont {Crosser}, \citenamefont {Leyden}, \citenamefont {Qi},\ and\
  \citenamefont {Minot}}]{Brown2016}%
  \BibitemOpen
  \bibfield  {author} {\bibinfo {author} {\bibfnamefont {M.~A.}\ \bibnamefont
  {Brown}}, \bibinfo {author} {\bibfnamefont {M.~S.}\ \bibnamefont {Crosser}},
  \bibinfo {author} {\bibfnamefont {M.~R.}\ \bibnamefont {Leyden}}, \bibinfo
  {author} {\bibfnamefont {Y.}~\bibnamefont {Qi}},\ and\ \bibinfo {author}
  {\bibfnamefont {E.~D.}\ \bibnamefont {Minot}},\ }\bibfield  {title} {\bibinfo
  {title} {Measurement of high carrier mobility in graphene in an aqueous
  electrolyte environment},\ }\href {https://doi.org/10.1063/1.4962141}
  {\bibfield  {journal} {\bibinfo  {journal} {Appl. Phys. Lett.}\ }\textbf
  {\bibinfo {volume} {109}},\ \bibinfo {pages} {093104} (\bibinfo {year}
  {2016})}\BibitemShut {NoStop}%
\bibitem [{\citenamefont {Gosling}\ \emph {et~al.}(2021)\citenamefont
  {Gosling}, \citenamefont {Makarovsky}, \citenamefont {Wang}, \citenamefont
  {Cottam}, \citenamefont {Greenaway}, \citenamefont {Patan{\`{e}}},
  \citenamefont {Wildman}, \citenamefont {Tuck}, \citenamefont {Turyanska},\
  and\ \citenamefont {Fromhold}}]{Gosling2021}%
  \BibitemOpen
  \bibfield  {author} {\bibinfo {author} {\bibfnamefont {J.~H.}\ \bibnamefont
  {Gosling}}, \bibinfo {author} {\bibfnamefont {O.}~\bibnamefont {Makarovsky}},
  \bibinfo {author} {\bibfnamefont {F.}~\bibnamefont {Wang}}, \bibinfo {author}
  {\bibfnamefont {N.~D.}\ \bibnamefont {Cottam}}, \bibinfo {author}
  {\bibfnamefont {M.~T.}\ \bibnamefont {Greenaway}}, \bibinfo {author}
  {\bibfnamefont {A.}~\bibnamefont {Patan{\`{e}}}}, \bibinfo {author}
  {\bibfnamefont {R.~D.}\ \bibnamefont {Wildman}}, \bibinfo {author}
  {\bibfnamefont {C.~J.}\ \bibnamefont {Tuck}}, \bibinfo {author}
  {\bibfnamefont {L.}~\bibnamefont {Turyanska}},\ and\ \bibinfo {author}
  {\bibfnamefont {T.~M.}\ \bibnamefont {Fromhold}},\ }\bibfield  {title}
  {\bibinfo {title} {Universal mobility characteristics of graphene originating
  from charge scattering by ionised impurities},\ }\href
  {https://doi.org/10.1038/s42005-021-00518-2} {\bibfield  {journal} {\bibinfo
  {journal} {Communications Physics}\ }\textbf {\bibinfo {volume} {4}},\
  \bibinfo {pages} {30} (\bibinfo {year} {2021})}\BibitemShut {NoStop}%
\bibitem [{\citenamefont {Hirai}\ \emph {et~al.}(2014)\citenamefont {Hirai},
  \citenamefont {Tsuchiya}, \citenamefont {Kamakura}, \citenamefont {Mori},\
  and\ \citenamefont {Ogawa}}]{Hirai2014}%
  \BibitemOpen
  \bibfield  {author} {\bibinfo {author} {\bibfnamefont {H.}~\bibnamefont
  {Hirai}}, \bibinfo {author} {\bibfnamefont {H.}~\bibnamefont {Tsuchiya}},
  \bibinfo {author} {\bibfnamefont {Y.}~\bibnamefont {Kamakura}}, \bibinfo
  {author} {\bibfnamefont {N.}~\bibnamefont {Mori}},\ and\ \bibinfo {author}
  {\bibfnamefont {M.}~\bibnamefont {Ogawa}},\ }\bibfield  {title} {\bibinfo
  {title} {Electron mobility calculation for graphene on substrates},\ }\href
  {https://doi.org/10.1063/1.4893650} {\bibfield  {journal} {\bibinfo
  {journal} {J. Appl. Phys.}\ }\textbf {\bibinfo {volume} {116}},\ \bibinfo
  {pages} {083703} (\bibinfo {year} {2014})}\BibitemShut {NoStop}%
\bibitem [{\citenamefont {Banszerus}\ \emph {et~al.}(2015)\citenamefont
  {Banszerus}, \citenamefont {Schmitz}, \citenamefont {Engels}, \citenamefont
  {Dauber}, \citenamefont {Oellers}, \citenamefont {Haupt}, \citenamefont
  {Watanabe}, \citenamefont {Taniguchi}, \citenamefont {Beschoten},\ and\
  \citenamefont {Stampfer}}]{Banszerus2014}%
  \BibitemOpen
  \bibfield  {author} {\bibinfo {author} {\bibfnamefont {L.}~\bibnamefont
  {Banszerus}}, \bibinfo {author} {\bibfnamefont {M.}~\bibnamefont {Schmitz}},
  \bibinfo {author} {\bibfnamefont {S.}~\bibnamefont {Engels}}, \bibinfo
  {author} {\bibfnamefont {J.}~\bibnamefont {Dauber}}, \bibinfo {author}
  {\bibfnamefont {M.}~\bibnamefont {Oellers}}, \bibinfo {author} {\bibfnamefont
  {F.}~\bibnamefont {Haupt}}, \bibinfo {author} {\bibfnamefont
  {K.}~\bibnamefont {Watanabe}}, \bibinfo {author} {\bibfnamefont
  {T.}~\bibnamefont {Taniguchi}}, \bibinfo {author} {\bibfnamefont
  {B.}~\bibnamefont {Beschoten}},\ and\ \bibinfo {author} {\bibfnamefont
  {C.}~\bibnamefont {Stampfer}},\ }\bibfield  {title} {\bibinfo {title}
  {Ultrahigh-mobility graphene devices from chemical vapor deposition on
  reusable copper},\ }\href {https://doi.org/10.1126/sciadv.1500222} {\bibfield
   {journal} {\bibinfo  {journal} {Sci. Adv.}\ }\textbf {\bibinfo {volume}
  {1}},\ \bibinfo {pages} {e1500222} (\bibinfo {year} {2015})}\BibitemShut
  {NoStop}%
\end{thebibliography}%
\end{document}